\documentclass[10pt,prl,aps,email,showpacs,amsmath,amssymb,twocolumn,superscriptaddress,nofootinbib,tighten,thanks]{revtex4-2}

\usepackage{graphics}
\usepackage{caption}
\usepackage{subcaption}

\usepackage[export]{adjustbox}
\usepackage{wrapfig}
\usepackage{babel,blindtext}
\usepackage{graphicx}

\emergencystretch=\maxdimen
\usepackage{hyperref}
\hypersetup{
colorlinks=true,
linkcolor=Blue,
filecolor=Magenta, 
urlcolor=Orange,
           }
\usepackage{color}
\usepackage{epstopdf}
\usepackage[dvipsnames]{xcolor}
\usepackage{dcolumn}
\usepackage{bm}

\usepackage{orcidlink}

\def\bea{\begin{eqnarray}} \def\eea{\end{eqnarray}}
\def\be{\begin{equation}}
\def\ee{\end{equation}}

\def\hl{}

\begin{document}

\title{Asymptotic Freedom and Vacuum Polarization Determine the Astrophysical End State of Relativistic Gravitational Collapse: Quark--Gluon Plasma Star Instead of Black Hole } 


\newcommand{\orcidauthorA}{0000-0002-9135-1396} 
\newcommand{\orcidauthorB}{0000-0003-4581-6313} 
\newcommand{\orcidauthorC}{0000-0002-6785-783X} 
\newcommand{\orcidauthorD}{0000-0001-8297-9336} 

\author{Herman J. Mosquera Cuesta\,\orcidlink{0000-0002-9135-1396}} 
\affiliation{Colciencias  
 National Program for Basic Sciences, Space Science, Avenida Calle 26, No. 57-41 Torre 8, Pisos 2-6, Bogota 111321, 
 Colombia } 
 \email{herman@icra.it}

 \affiliation{Valencia International University, Astronomy and Astrophysics Program, C/ Pintor Sorolla, 21, 46002 Valencia, Spain}
 
\author{Fabián H. \hl{Zuluaga Giraldo}\,\orcidlink{0000-0003-4581-6313}}
\affiliation{Escuela de Fisica, Universidad Nacional de Colombia, Campus Medellin, A.A. 3840, Medellin, Colombia}
\email{fhzuluagag@unal.edu.co} 

\author{Wilmer D. Alfonso Pardo\,\orcidlink{0000-0002-6785-783X}} 
\email{wdalfons@unal.edu.co}
\affiliation{Instituto de Fisica, Universidad de Antioquia, Campus Medellin, A.A. 1226, Medellin, Colombia}
 
\author{Edgardo Marbello Santrich} 
\email{ejmarbellos@unal.edu.co}
\affiliation{Escuela de Fisica, Universidad Nacional de Colombia, Campus Medellin, A.A. 3840, Medellin, Colombia}

\author{Guillermo U. Avendaño Franco\,\orcidlink {0000-0001-8297-9336}}
\email{gufranco@mail.wvu.edu }
\affiliation{Department of Physics and Astronomy, West Virginia University, P.O. Box 6201, Morgantown, WV 26506-6315, USA}  

\author{Rafael Fragozo Larrazabal} 
\email{rfragozolarrazabal@alumnos.viu.es} 
\affiliation{Valencia International University, Astronomy and Astrophysics Program, C/ Pintor Sorolla, 21, 46002 Valencia, Spain}

\begin{abstract}
A general relativistic model of an astrophysical hypermassive extremely magnetized ultra-compact self-bound quark--gluon plasma (QGP: ALICE/LHC) object that is supported against its ultimate gravitational implosion by the simultaneous action of the vacuum polarization driven by nonlinear electrodynamics (NLED: ATLAS/LHC: light-by-light scattering)---the vacuum ``awakening''---and the asymptotic freedom, a key feature of quantum chromodynamics (QCD), is presented.  
These QCD stars can be the 
final figures of the equilibrium of collapsing stellar cores permeated by magnetic fields with strengths well beyond the Schwinger threshold due to being self-bound,
and for which post-supernova fallback material pushes the nascent remnant beyond its stability, forcing it to collapse into a hybrid hypermassive neutron star (HHMNS). Hypercritical accretion can drive its innermost core to spontaneously break away color confinement, powering a first-order hadron-to-quark phase transition to a sea of ever-freer quarks and gluons. This core is hydro-stabilized by the steady, endlessly compression-admitting asymptotic freedom state, possibly via gluon-mediated enduring exchange of color charge among bound states, e.g., the odderon: a glueball state of three gluons,  
or either quark-pairing (color superconductivity) or tetraquark/pentaquark states (LHCb Coll.). 
This fast---at the QGP speed of sound---but incremental quark--gluon deconfinement  unbinds the HHMNS's baryons so  catastrophically that   transforms it, turning it inside-out, into a neat self-bound  QGP star. 
A solution to the nonlinear Tolman--Oppenheimer--Volkoff (TOV) equation is obtained---that clarifies the nonlinear effects of both NLED and QCD on the compact object's structure---which clearly indicates the occurrence of hypermassive QGP/QCD stars with a wide mass spectrum ($0\lesssim$ M$^{\rm{QGP}}_{\rm{Star}}\lesssim$\,7\,M$_\odot$ and beyond), for star radii  ($0\lesssim R^{\rm{QGP}}_{\rm{Star}}\lesssim 24$\,km and beyond) with B-fields ($10^{14} \leq$ B$^{\rm{QGP}}_{\rm{Star}} \leq 10^{16}$\,G and beyond). 
This unexpected feature is described by a novel mass vs. radius relation  derived within this scenario. Hence, endowed with these physical and astrophysical characteristics, such QCD stars can  definitively emulate what the true (theoretical) black holes are supposed to gravitationally do in most astrophysical settings. 
This color quark star could be found through a search for its eternal ``yo-yo'' state gravitational-wave emission, or via lensing phenomena like a gravitational rainbow (quantum mechanics and gravity interaction), 
as in this scenario, it is expected that the light deflection angle---directly influenced by the larger effective mass/radius (M$^{\rm{QGP}}_{\rm{Star}}(B)$, R$^{\rm{QGP}}_{\rm{Star}}(B)$) and magnetic field of the deflecting object---increases as the incidence angle decreases, in view of the lower values of the impact parameter. 
The gigantic---but not infinite---surface gravitational redshift, due to NLED photon acceleration, makes the object appear dark. 
\\
This arXiv.org article is a version of the original article published by the MDPI Publishing Co. journal \textit{Universe} at the link \url{https://www.mdpi.com/2218-1997/11/11/375}
\end{abstract}

\thanks{Corresponding author: herman@icra.it}




\date{\today}
\maketitle

\section{Introduction}
\label{introd}

The concept of a black hole has been an essential ingredient of our modern view of the universe through relativistic astrophysics and in cosmology as a sort of cosmic seed for galaxy formation and dynamics. The~first interferometer-detected gravitational wave signal had its source interpreted as a merged stellar-mass black hole binary (LIGO 2015) \cite{LIGO-VIRGO(2015)}.    Regarding black holes, last year Roy P. Kerr, who discovered the metric field describing a rotating gravitational source, raised questions about the reality of their singularities and of the Hawking--Penrose singularity theorems while asserting that Kerr metric indeed describes a non-singular astrophysical ultra-compact~star \cite{Kerr-No-Singularities(2023)}.

These backgrounds, together with the LHC momentous achievements---quark/gluon plasma (QGP, ALICE) and light scattering off light (ATLAS)---are strong motivations for revising the general relativistic ultimate profiles of collapsing configurations of extremely magnetized ultra-compact stellar cores within the framework of quantum chromodynamics and nonlinear electrodynamics. 

From now on, we shall refer to such objects as self-bound quark--gluon plasma (QGP) stars. Self-bound means that, in~theory, the~star is held together by the strong nuclear force that binds quarks and gluons, rather than by gravity alone, which is weaker, so the star does not rely solely on the gravitational pull to maintain its structure. A~color quark star could exist in equilibrium with a vacuum and have a sharp, abrupt edge where its density and pressure fall to zero. A~thin crust of normal matter (neutrons, protons, and~electrons) may also be present, though, by~virtue of the remnants of the deconfinement phase transition. Self-boundedness is a characteristic that enhances the QGP star's intrinsic magnetic field to strengths higher than $10^{18}$\,G at \linebreak its surface. 

{
Meanwhile, there are a  number of astrophysical environments where extremely magnetized hypermassive QGP stars can be formed through hypercritical accretion. They include the  gravitational collapse-at-once of  super strongly magnetized and supermassive main-sequence or evolved/supergiant branch stars (e.g., WR HD45166, Antares, Betelgeuse, etc.), the~core-collapse supernovae-producing hypermassive neutron star pulsars (HMNSPs) that can subsequently accrete large amounts of fallback material to collapse further; the coalescence and merger of binaries constituted by  evolved high-mass stars orbited by  equally massive B-type stars (e.g., WR 104), the~coalescing and merging compact binary systems having highly magnetized (e.g., Swift J0243.6+6124) massive or supermassive neutron stars/pulsars (the multi-messenger sources), and~also the high-mass X-ray binary systems where the primary companion is a massive or supermassive neutron star orbiting a main-sequence high mass or an evolved massive Roche-overflowing secondary star. Also, collapsars can be included as potential sites.}

Post-supernova ($\sim$10\%) fallback material pushes the proto-neutron star over its stability, causing it to collapse into a hybrid hypermassive neutron star (HHMNS). Ulterior accretion can drive its innermost core to spontaneously break away color confinement, powering a first-order hadron $\rightarrow\rightarrow$ quark phase transition to a sea of ever-freer quarks and gluons zipping about at relativistic speeds. 
{This core is hydro-stabilized by the steady,} endlessly compression-admitting asymptotic freedom, possibly via gluon-mediated enduring exchange of color charge among bound states, e.g.,~the odderon: a glueball state of three gluons, or~the lightest pseudo-scalar glueball state X(2370) discovered in (2011) and that had its nature confirmed in (2024) at BESIII. 
{Inside a QGP star, the QCD-``feeling'' particles can also coalesce to form the color-superconducting phase (quark-pairing) or the tetraquark/pentaquark bound states as discovered by LHCb Collaboration (2022). }

Standardly, what determines the QGP star's final fate is the initial density and velocity profile of the collapsing shells. This approach overlooks nonlinear effects on the core's structure due to both the  positive potential energy of QCD asymptotic freedom and NLED quantum vacuum polarization that kick in at field strengths beyond the Schwinger limit. This core gravitationally survives  eternally on the brink of ultimate collapse---the GECKO state---preventing the catastrophic implosion to the would-be singularity. This way, the QGP star radius R$^{\scriptscriptstyle{\rm  GECKO}}_{_{\scriptscriptstyle{\textrm{QGP Star}}}}$ $\gtrsim$\ and the 2M $>$ Schwarzschild radius for any given mass are defined. 
Further, photon acceleration triggers a non-infinite surface gravitational redshift $z_{_{\rm Grav}}\gtrsim10^8$, making the QGP star to look like a dark body. 

We construe the whole picture given above as though nature would ''prefer'' an exceptionally happening nonpareil explosion---in the GECKO's attempt to trap as much latent heat via gravity as the energy released at the spontaneous breaking of GUT gauge symmetry during the primeval universe---rather than a breach in~space--time.

{
The general outline of this paper reads as follows: The introductory section just given is supplemented and fused together with some key additional arguments supported by the results of particle accelerators like LHC and RHIC. Section~\ref{sec2} offers an astrophysical motivation built around the concept of black hole in contemporary astrophysics. In~particular, it discusses the astrophysical ultimate state of relativistic gravitational collapse, the~QGP/QCD star in our model rather a singular compact object; the so-called true (theoretical) black hole. Such a section delivers a direct confrontation to the purported black hole picture as the astrophysical compact body said to reside and lurk in most galactic and extragalactic astronomical sources.
Next, another subsection offers a clarifying discussion on the different mechanisms acting on the stellar remnant with regard to the  magnification of the core's magnetic field upon gravitational collapse to a typical neutron star vs. collapse to a color-superconducting quark core.
This section is substantiated by a short account on the momentous results achieved by the ATLAS/ALICE LHC particle accelerator (see Figures \ref{Figure 3} and \ref{Figure 1}). Those specific results become the guiding ideas to explore the concept of extremely magnetized hypermassive quark--gluon plasma star that we introduce in this paper.
The core of the paper, Section~\ref{sec:5}, directly  
develops the proper QGP star model by elaborating
the derivation of the nonlinear Tolman--Oppenheimer--Volkoff (TOV) equation and then  discussing about both its solutions and the resulting and novel QCD star mass vs. radius relations, and~other related thermodynamic properties of it, derived within this framework.  
In closing this section,  we analyze and interpret the novel equation upon which to build the M vs. R plot that illustrates the main result obtained within this context, exhibiting a wide mass spectrum (see Figure
~\ref{Fig.-4} below): 

\vspace{-12pt}
\begin{widetext}
\centering 
\begingroup
\makeatletter\def\f@size{9}\check@mathfonts
\def\maketag@@@#1{\hbox{\m@th\normalsize\normalfont#1}}
\be
{\rm{Mass:}} \, 0\lesssim M^{\rm{QGP}}_{\rm{Star}}\lesssim\,7\,M_\odot \longrightarrow, {\rm{\,radii:}} \, 0\lesssim R^{\rm{QGP}}_{\rm{Star}}\lesssim 24\,km \, \longrightarrow, {\rm{\, fields:}} \, 10^{14} \leq B^{\rm{QGP}}_{\rm{Star}} \leq 10^{16}\,G \, \longrightarrow.
\ee
\endgroup
\end{widetext}

In 
the closing remarks section that summarizes  the main body of this paper, some conclusions are drawn by addressing the formal implications of the relationship between mass and radius derived here for current relativistic astrophysics research as well as for our overall understanding of the universe.
The remainder of the paper highlights the essential physical and astrophysical components of this QGP/QCD star scenario that are not duly elaborated in the depicted central body of the article.

\section{On the Astrophysical Ultimate State of Relativistic Gravitational Collapse: A QGP/QCD~Star}
\label{sec2}

\subsection{Astrophysical~Insights} 

Soon after having presented his theory of general relativity (GR), Einstein figured out the existence of travelling space--time---the gravitational metric field ($g_{\mu\nu}$)---curvature waves propagating at the vacuum speed of light, the~gravitational radiation (GW). Among~plenty of potential cosmic sources, it can be produced by accelerating very massive and ultra-compact astrophysical bodies like coalescing and merging neutron star and/or black hole (BH) binaries.

Supplementing the decades-long follow-up of the orbital dynamics of the PSR\,1913+16 binary pulsar via radio telescopes by Taylor and Hulse~\cite{PSR1913+16_Follow-up}, the~event GW\,150914~\cite{LIGO-VIRGO(2015)}---identified by LIGO Observatories in 2015---confirmed the reality of GW by the direct measurement of the deformation---on the atomic nucleus size scale---of the physical structure of Fabry--P\'erot Michelson interferometers. This signal came from a source at a luminosity distance of $\sim$1.3\,Gly, which was interpreted as a binary of bare black holes of stellar mass. By~bare, we mean that, according to a handful of astronomical observatories, neither the hallmark of accretion discs nor the signatures of astrophysical jets were identified from this signal of the binary collision. Likewise, neither hints of magnetic fields nor the presence of concomitant electromagnetic emissions emerged from the event. Moreover, no imprint of astro-particle radiation, e.g.,~neutrino bursts, or~any other astrophysical characteristic, was observed in such a signal. See the LIGO homepage: 
\url{https://www.ligo.org/detections/GW150914.php} (accessed on 15 August 2024) and the announcement paper for the event~\cite{LIGO-VIRGO(2015)}. This is contrary to what did happen with the spectacular GW170817 global astronomy multi-messenger event, wherein most of those astrophysical characteristics were detected by a number of space- and ground-based observatories. See LIGO homepage: \url{https://www.ligo.org/detections/GW170817.php}  (accessed on 15 August 2024) and Nakar's review and references therein focused on the signal electromagnetic counterparts~\cite{GW170817-EM-Counterparts(2020), nakar(2020)}.

In connection with BH physics and astrophysics, last year Roy Kerr, the~theorist who developed the space--time metric for the vacuum exterior to a rotating astrophysical body, raised questions on the reality of BH singularities and the Penrose--Hawking singularity theorems~\cite{Kerr-No-Singularities(2023)}. Among~the crucial ideas brought up for discussion, Kerr asserted that ``non-singular collapsed neutron stars can generate Kerr (metric)'', and~further stated that ``these solutions are merely substitutes for a non-singular interior star with a finite boundary at or inside the (Kerr metric) inner horizon.'' The debate rages~on.

\subsection{Flux Conservation During Collapse to Typical Neutron Stars vs. Field Amplification of Stellar Cores Collapsing to Color-Superconducting Quark~Cores} 

\subsubsection{Virial Theorem and Collapse to Canonical Neutron~Stars}

{An exception was found in the neutron stars known as magnetars
~\cite{magnetars-field}, 
whose magnetic fields reach $B_{\rm Mgns}\simeq 10^{15-16}$\,Gauss (G) \cite{Paret-etal(2020)}; no other astrophysical compact stars are known to have field strengths higher.
Indeed,  within~a newly formed neutron star, the rapid rotation and convective motions can act as a dynamo, a~mechanism that amplifies magnetic fields. The~rotation stretches and twists existing magnetic field lines, while convection, driven by temperature gradients, further stirs and organizes these fields, leading to a significant amplification. Specifically, in~nascent neutron stars, the intense fluid dynamics and powerful magneto-rotational instabilities, such as the Kelvin--Helmholtz shear instability, create the conditions for a small-scale dynamo to operate, rapidly amplifying the pre-existing or inherited magnetic field. As~an example, the~Tayler--Spruit dynamo is a specific model that suggests that magnetic fields are amplified in a highly turbulent, convective, and~differentially rotating proto-neutron star interior. Processes like these are leading candidates for creating some of the most powerful currently known fields, with~strengths reaching $10^{16}$ Gauss, those of magnetars.
In connection to this last point, decades ago, Lerche and Schramm~\cite{Lerche-Schramm(1977)} precluded the occurrence of astrophysical objects with $B_{e^\pm} \simeq 10^{20}$\,G whenever both dynamical (some sort of breaking) and quantum-mechanical effects (e$^\pm$ pair production) were taken into account. This constrains such a field to $B \sim 10^{16}$\,G, a~strength similar to those found in  magnetars~\cite{magnetars-field}. However, such an analysis is restricted to compact objects constituted mainly by nuclear matter, but~not to (color-superconducting) {quark-matter cores. }
}

\subsubsection{Field Amplification During HHMNS Collapse to Color-Superconducting Quark~Cores}

{Theoretical estimates for magnetic field strengths of color-superconducting quark cores range from 1\,GeV$^2$/$10^{15}$\,G to $10^3$\,GeV$^2$/$10^{18}$\,G. Specifically,  the~rotation of quarks around an internal  color-magnetic field arising from color superconductivity (CS)  can also contribute to the overall electromagnetic field (the color superconductor is an electromagnetic insulator; thus, it preserves fields that  penetrate the CS phase).
These high fields are thought to be spontaneously generated by the quark matter itself---likely via the appearance of a BEC state, as~quoted in footnote 2, below---and are influenced by the color-superconducting phase through mechanisms like the Savvidy mechanism (1993). It refers to the Savvidy vacuum in quantum chromodynamics (QCD), which describes the behavior of pure Yang--Mills theory in a background magnetic field. It was initially calculated using a one-loop approximation } \cite{Savvidy mechanism}.  
The theory studies the effective potential of a three-dimensional lattice gauge theory in the presence of an external chromo-magnetic field. The~calculations support Savvidy's ``ferromagnetic'' vacuum picture, where the vacuum state becomes non-trivial in the presence of a super strong magnetic field. 

{On this basis, a~number of current theoretical studies have indicated that stellar cores collapsing to color-superconducting quark cores, as~remnants of massive stars (10--25\,M$_\odot$), can build surface magnetic fields up to $B_{\rm Surf} \sim 10^{17}$\,G, while their inner strengths may range from 10$^{18}$\,G for nuclear matter~\cite{Broderick-etal(2000), Glendenning book, Weber book} to 10$^{20}$\,G for quark matter,  because~of its self-bound feature~\cite{EF-Incera-etal(2019), Ferr-etal(2010), ferrer-etal(2015), papers-NS-deformation-1, IJMPD(2007)}.  
In fact, Sotani and Tatsumi estimated that the critical magnetic field strength at which the lowest and second Landau levels play an important role in the quark phase inside massive (M $\simeq 2.8$\,M$_\odot$) hybrid stellar cores should be $B \sim 10^{19}$\,G at the nuclear density $\sim 0.16 \,{\rm fm}^{-3}$ \cite{Sotani-Tatsumi(2015)}. Moreover, last year, Fraga, Palhares, and~Restrepo studied the case of symmetric quark matter considering magnetic fields in the range $eB \sim (1-9)$\,GeV$^2$ ($B \sim 1.7 \times 10^{20}-1.53 \times 10^{21}$)\,G} \cite{Fraga-etal(2023)}. 

{
Aside from the previous arguments, experimental results in relativistic heavy-ion collisions (RHIC/USA) are shown to have produced hypercritical (well beyond the Schwinger limit: $\rm B_{Sch} \simeq 10^{13.5}$\,G $\leftrightarrow E_c = \frac{m^2_e c^3} {e\hbar} \simeq 10^{16}$\,V\,cm$^{-1}$) magnetic fields on the order of $B \sim 10^{19}$\,G } \cite{STAR-RHIC(2010), adam-etal(2020), aboona-etal(023)}. 
{And finally, the~current highest B field upper value was set from the ATLAS/LHC CERN data after that experiment evidenced  the presence of the QGP state (2017): $B_{c} =  {\rm B^{B-I}_{ATLAS} } \simeq 90\,\rm{GeV^2} \simeq \,(3.28) \cdot 10^{22}$\,G for NLED a l\'a Born-Infeld }
\cite{Ellis-etal.(2017)}. It is the sort of strength called for in the present paper to support our model of a hypermassive, ultra-strongly magnetized QGP/QCD star, a~class of astrophysical objects that has yet to be discovered. (Recall that the critical ($_{ct}$) magnetic field: $B_{ct}^{^f} \equiv (m^2_{_f} c^2) / (e_{_f} \hbar)$ depends on the mass ($m_{_f}$) and charge ($e_{_f}$) of the specific quantum field ($f$)). 
{In what follows, we take the above backgrounds as sound evidence, as~well as other essential pieces of galactic and extragalactic BH astrophysics, including those discussed after the Event Horizon Telescope Network observations on M87 and Sgr A$^\star$ \cite{EHT-BH-Radio-Images}. } 

Thus being, in~the present work we explore the implications of a general relativistic model of ultra-compact quark-core remnants that are pervaded by an extremely strong magnetic field with strengths well beyond the Schwinger limit that are described in the framework of Born--Infeld nonlinear electrodynamics (NLED)---a long-standing theory that was recently validated by ATLAS/LHC (2017) \cite{ATLAS/LHC-Coll.(2017), ATLAS:ligh-by-light}. The~hypermassive nuclear-matter core can  undergo, upon~hypercritical accretion, a~catastrophic transition from nucleonic/hadronic matter to deconfined quark matter.   Specifically, the~strange matter is energetically favorable to neutron matter and light quark matter. Hence, the~whole HHMNS can be converted into the sort of ultra-compact object here proposed; a strange QGP star, based on what recent ALICE/LHC experiments indicate~\cite{ATLAS/LHC-Coll.(2017)}. See \mbox{Figures \ref{Figure 3} and \ref{Figure 1}.}


\begin{figure}[t!]
\includegraphics[width=8.0cm
]{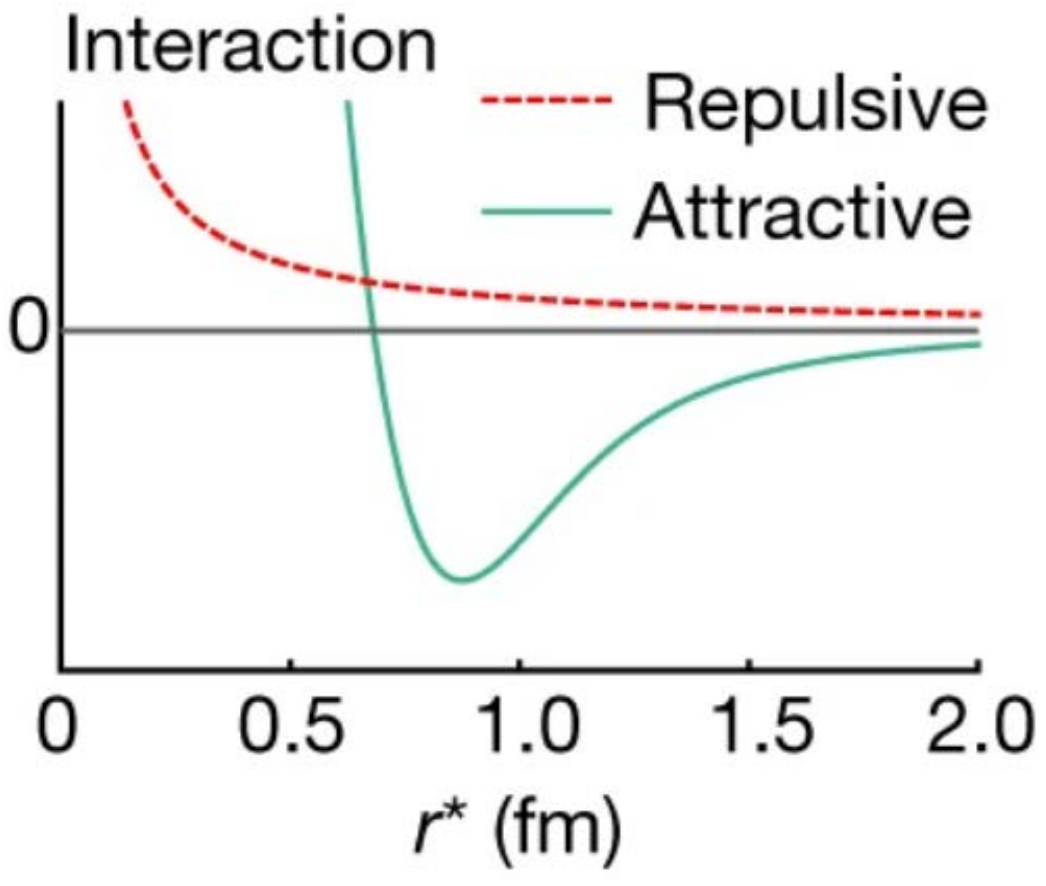}
\caption{Plot 
 of the attractive (green curve) vs. repulsive (red curve) potentials $V(r)$ as a function of distance $r$, for~the QCD interaction. (Plot taken from ''Unveiling the strong interaction among hadrons at the LHC'' \cite{ALICE(2020)}). } 
\label{Figure 3}
\end{figure}

\subsection{QED Vacuum Polarization/QCD Asymptotic~Freedom}

The announcement on 4 July 2012 on the discovery of the Higgs boson~\cite{ALICE(2024), Higgs-Boson(2012)}, was supplemented by the observation of the phenomenon of light-by-light scattering achieved by ATLAS/LHC$|$CERN (2017) by studying the quark--gluon plasma (QGP) produced at ALICE/LHC from lead-ion collisions (Pb$^{82}$+Pb$^{82}$:\,$\sqrt{\rm S_{_{NN}}}$= 5.02\,TeV), in~evidencing it at a 8.2\,$\sigma$ certainty level.  (The QGP is a deconfined state of matter in which interactions binding the quarks and gluons---components of nucleons: protons/neutrons---do not occur anymore, which is the QCD asymptotic freedom property at work.) This state fragments out through highly collimated photon-tagged jets that dump their energy through the phenomenon dubbed jet quenching,  
a~sort of hadronization process~\cite{ATLAS/LHC-Coll.(2017), Kate-Scholberg(2021)}. Bear in mind, en passant, that proton--proton collisions are not energetic enough to produce a quark--gluon plasma. That is why ALICE had to wait a while, up~to 2015, for~the lead--lead ($Pb+Pb$) ion collisions to come alive, as~they do produce~QGP. 

All of these measurements of hadron interactions at ALICE have revealed novel features that have broad implications for nuclear physics and relativistic astrophysics. (See the review by the ALICE Collaboration on the long way through QCD with the ALICE experiment~\cite{ATLAS/LHC-Coll.(2017)}). Specifically, ALICE indicates that interactions between a proton (p) and a Lambda (${\Lambda}$), Xi (${\Xi}$), or~Omega (${\Omega}$) hyperon---unstable particles containing strange \mbox{quarks---are} attractive. In~relativistic astrophysics, it is foreseen that these interactions may play a role with respect to the stability of a HHMNS (${\rm 2\,\leq M^{HHM}_{NS}({M_\odot}) \simeq 5}$) with a large radius. (See~\cite{Maximum-NS-Mass_Kalogera-Baym(1996)} for a theoretical limit on the maximum NS mass: ${\rm{M_{NS}\,\lesssim\,3\,M_\odot}}$, which is valid for up to twice the nuclear-matter saturation density.)

\begin{figure}[t!]
\includegraphics[width=8.0cm]{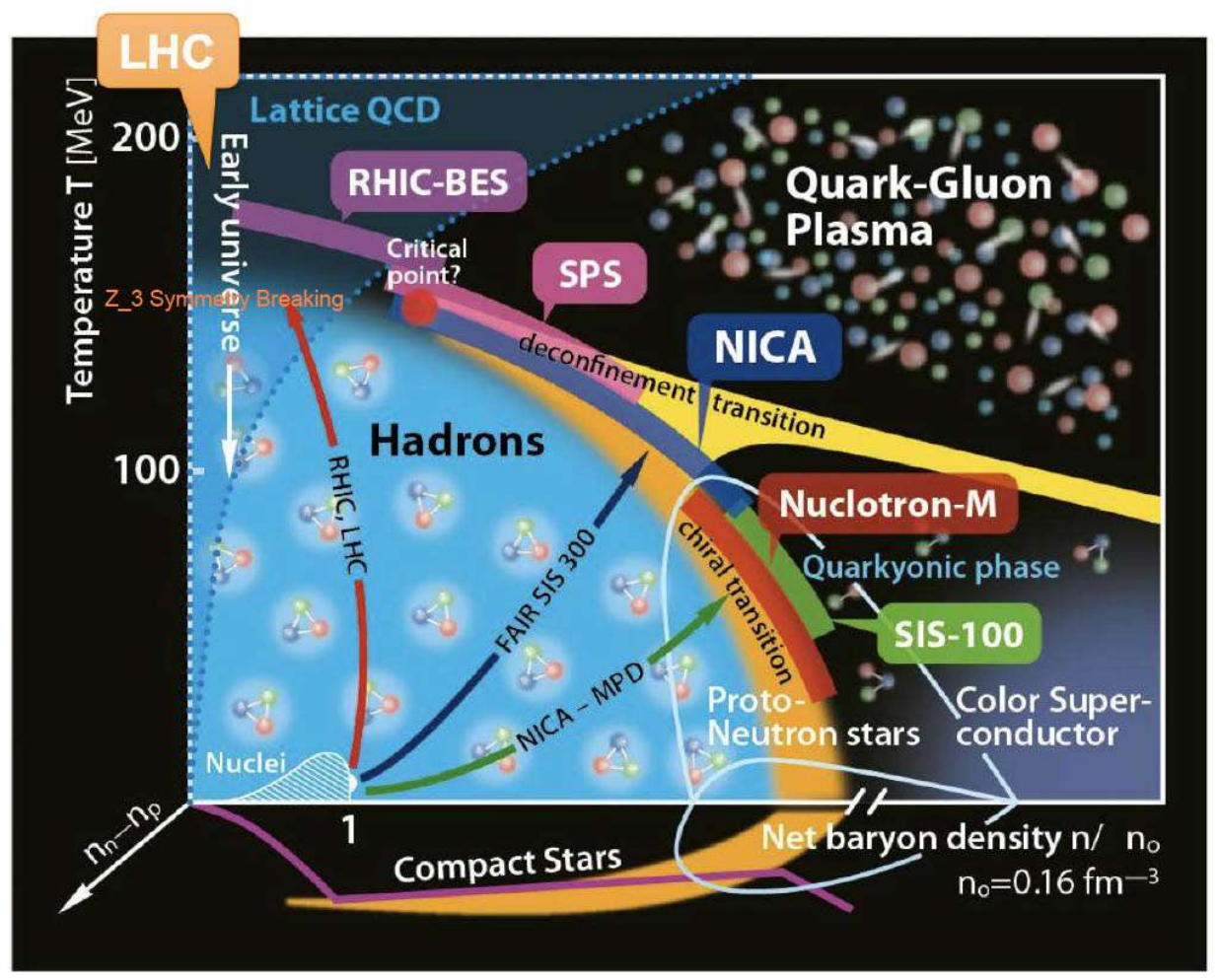}
\caption{This plot highlights the location of proto-neutron stars, the~deconfinement transition, the~QGP state as per heavy-ion collisions, and~the Lattice QCD limit. (Plot taken from ``Introduction to Heavy-Ion Physics|ATLAS Open Data''.  
 \cite{ATLAS-open-data})
The Z$_3$-symmetry breaking label was introduced by \linebreak the authors.}
\label{Figure 1} 
\end{figure}



Such a momentous result for ATLAS allows for the identification of the self-interacting propagation of photons as predicted by nonlinear electrodynamics (NLED) \cite{Euler-Kockel(1935), Heisenberg-Euler(1936), plebansky_lectures_NLED, Born-Infeld(1934)}. 
In~fact, NLED was the very first step in building the quantum theory of electron and photon fields, the~QED~\cite{Feynman(QED-1985)}. Not surprisingly, photon--photon self-interactions cannot be accounted for within classical Maxwell's electrodynamics, as~light cannot interact with itself by virtue of being electrically neutral, that is; the electromagnetic (E-M) field carries no charge, i.e.; the E-M has no self-coupling
\footnote{{Math/ly: linear in Maxwell scalar $F$ = $F_{\mu\nu}F^{\mu\nu}$, $\therefore$ $F_{\mu\nu}(A_\mu) = \partial_\mu A_\nu - \partial_\nu A_\mu$: the anti-symmetric electromagnetic field tensor; $\therefore$ $A_\nu$\,=\,$(\phi/c, \vec{A})$: the electromagnetic 4-potential; $\phi$: the electric scalar potential; $c$: the vacuum speed of light $\therefore$ $\vec{A}$: the magnetic vector potential.}}. Such a feature is contrary to what is predicted, for~instance, by~the non-abelian Yang--Mills gauge field theory describing the weak and strong nuclear interactions that undergo~self-coupling.

In the accelerator environment, most composites decay quite rapidly, and~the QGP state may last for $\Delta T^{\rm QGP}_{\rm decay}$\,$\sim$\,$10^{-23}$\,[s] as its fireball expands, cools, and~transforms into hadronic showers (jets) of particles. However, in~an accretion-driven neutron star undergoing a catastrophic color-charge deconfinement, i.e.,~inside a hydro-stabilized self-gravitating QGP star, there is no such way to decay back because the extremely high pressure, an~injection of gigantic amounts of energy from its overwhelming self-gravitation strength, must rein \linebreak it in\footnote{A self-bound QGP star is a sort of astrophysical self-magnetization state of  electrically charged quarks as a consequence of Bose--Einstein condensation (BEC) of such particles driven by gravitational shrinkage and color-charge deconfinement followed by quark-pairing (color superconductivity). The~major role played by BECs not only in condensed matter physics but also in relativistic astrophysics was explained in detail a few years ago by Chavanis in~\cite{Chavanis(2015)}, by~highlighting the dynamical effect of the bosons' repulsive scattering, i.e.,~the Heisenberg uncertainty principle's degeneracy negative pressure, in~stabilising galactic structures such as dark matter halos and astrophysical compact objects against gravitational collapse. Furthermore, ref.~\cite{Chavanis(2015)}  emphasizes the fundamental role of the bosons' scattering length in dictating the sort of collapsed remnant to wait for. In~the case of stable boson stars featuring a positive scattering length, they could mimic the purported supermassive black holes said to reside at the core of most galaxies and other similar astrophysical ultra-compact galactic sources. See also the gravitational vacuum BEC alternative to black holes by Mazur and Mottola~\cite{Mazur-Mottola-GRAV_BEC_Object(2023)}. Aside from this, from~a theoretical point of view, these BEC states are no longer considered thermal states, so they cannot be assigned a temperature; i.e.,~as the substance is not in (or near) equilibrium, there is no temperature. The~preparation of a BEC as a deterministic variation of the quantum ground state is effectively a state of indeterminate temperature. 
This is simply an alternative framing of the Heisenberg Uncertainty Principle.}.


{In this understanding, such particles---quarks and gluons---can similarly coalesce to form bound states in the dense QGP matter in the interior of either HHMNSs or QGP stars, much as that which appears to occur in either the neutrons' super-fluid state $^3$P$_2$ which induces a super-strong magnetic moment and field strength due to the neutron's Cooper pairing, a~sort of color ferromagnetism in the QGP star's quark matter~\cite{Iwazaki(2005)}, or~in the universal effective distribution of quarks and gluons observed inside correlated nucleon pairs~\cite{quark-gluon-distribution(PRL2024)}. This state can undergo a sort of exchange among all of the species inside the QGP star matter while it struggles between the superpower gravitational strength and the total repulsive pressure in the yo-yo state described above. This coalescence process is quite similar to what is envisioned for the other higher QCD resonances in particle accelerators like RHIC~\cite{STAR-RHIC(2010)}. However, in~the present case, it is a result of the catastrophic rearrangement associated with color-charge deconfinement inside the gravitation-dominated transitioning HHMNS (see Figure \ref{Figure 1}).  
}

{Next, we construct the nonlinear TOV equation for the QGP star in the framework of nonlinear electrodynamics (NLED) supplemented by the physics of the quantum chromodynamics (QCD), especially by asymptotic freedom. Further arguments in support of this theoretical model stemming from different fields of physics: theoretical particle physics (QCD, QED), experimental particle and nuclear physics (LHC, RHIC), nuclear astrophysics, nonlinear electrodynamics (NLED), and~observational astrophysics are also given along the section. }

\section{Deriving Nonlinear TOV Pressure-Gradient Equation for Equilibrium~Configurations}
\label{sec:5}
The quasi-static evolution of the very central interior (core) of massive stars is described by the Tolman--Oppenheimer--Volkoff (TOV) equation~\cite{TOV,Oppenheimer-Snyder(1939), TOV-2}. In~what follows, we construct a spherical stellar model that will be able to avoid collapsing to the purported black hole. The~standard procedure to derive the TOV equation involves using Einstein's equations for a general time-invariant, radially self-similar metric, with~a shell mass of uniform/homogeneous density, i.e.,~there is no shell-crossing nor shock waves; thus, the model neglects the effects of backscattering.
This means that during gravitational collapse, the condition is realized by the metric function $\phi(r,t)$, while the radial velocity profile is ruled by
$u$=$-\dot{R}(m(r,B), t)$, whose end state, the isotropic, spherically symmetric ($t, r, \theta, \phi$) Schwarzschild space--time, is featured by the line-element---here, $R$ represents the circumferential radial coordinate. 
\be
ds^2 =  -\left[1-\frac{2 G_{_{\rm N}} m(r,B)}{r c^2}\right] c^2 dt^2\,+\,
\frac{dr^2}{\left[1 - \frac{2G_{_{\rm N}} m(r,B)}{r c^2} \right]} 
+ R^2 d\Omega^2. 
\label{Schwarzs-Metric}
\ee
Then, we incorporate quantum electrodynamics (QED) effects by resorting to the effective metric (Equation (\ref{NLED-eff-metric})) when computed for the Born--Infeld NLED Lagrangian so that the Schwarzschild-like line-element 
(Equation (\ref{Schwarzs-Metric})) transforms into~\cite{GRAV-REDSHIFT(2004), MNRAS(2004), IJMPA(2006)}

\begin{equation}
ds^2 = {f(L(\tilde{\mu}(\hbar),B)} \left\{- {e^{\phi(r,t)}}\,c^2 dt^2 + e^{\Psi(r,t)} {dr^2} + {R^2} d\Omega^2 \right\}, 
\label{eff-metric}
\end{equation}
where $d\Omega^2 = (d\theta^2 + \sin^2\theta d\phi^2)$ and the metric function $\phi(r,t)$ is determined by using Einstein's equations in connection to the null-divergence constraint on the matter, plus the nonlinear electromagnetic fields' energy--momentum--stress tensor featuring the eternally collapsing stellar core, the~GECKO. In~this metric, the~component $ e^{\Psi(r,t)} = {\left(1 - \frac{2 G_{_{\rm N} } m(r,B)} {rc^2} \right) }$ comes out as is shown~next.

In using the TOV equation to model the end state of an imploding bounded sphere of some material in a vacuum, both (i) the surface zero-pressure condition $p(r)\,$=\,0 indicating the core radius and (ii) the metric stationarity and continuity conditions $e^{\phi(r,t,B)}|_{_{r\leq R_{Schw}}} \,$=\,$1 - 2G_{_{\rm N}} M(r,B)|^{Star}_{QGP}/c^2r$ (see Equation~(\ref{mass}) below) should be imposed at the boundary to match the Schwarzschild metric Equation~(\ref{Schwarzs-Metric}) by the Birkhoff theorem. Recall that the proper time of collapse is computed as $\tau = \frac{\pi}{2} R_0 \left(\frac{R_0}{2GM} \right)^\frac 1 2$ $\therefore$ $R_0, M$ star radius and mass at collapse~starting.


Now, as~in standard stellar astrophysics, we define a new quantity $m(r,t,B) \doteq \frac{1}{2} r(1 - e^{-2\Psi(r,t,B)}) \Leftrightarrow g^{-1}_{rr}$, representing the mass inside a collapsing shell of radial coordinate $R$ in such a way that it satisfies $m(r$ = $0)$ = 0 (hereafter, $\rho^{Total}_{(r,B)} = \rho(r,B)$: mass density, $p(r,B)=p^{Total}_{(r,B)}$: pressure, in~the magnetized stellar fluid). Thus,
\begin{equation} 
\frac{dm(r,B)}{dr} = 4\pi \, R^2 \, \rho^{Total}_{(r,B)} = 4\pi \, R^2 {\rho(r,B) } \, .
\label{core-mass}
\end{equation}

Further, from~the Bianchi identities, it follows that the energy--momentum--stress tensor conservation condition~reads

\be
\underbrace{ \nabla_{\mu}  \langle T^{\mu\nu } \rangle = 0 }_{\textrm{En-Mom Conserv}} \, .
\label{T-conserv}
\ee

Then, after~projecting the above equation in the transversal direction to the fluid 4-velocity (turning the collapse dynamics description $t$-independent), and~dropping the ($r,B$)-dependence: $\phi(r,B)=\phi$, $\rho(r,B)=\rho$, $m(r,B)=m$, $p(r,B) = p$, and~recalling the perfect fluid energy--momentum tensor:
$T_{00} = - \rho c^2$ and $ T_{ij} =  p \delta_{ij}$,
one obtains the r-dependent metric field $\phi$ equation 

\be
\nabla_\nu T^{\nu}_r = 0 \rightarrow  {f\big(L(\tilde{\mu}(\hbar),B\big)}
\frac{d \phi}{dr} = - \frac{2}{\left({p} + \rho {c^2} \right)}  \left(\frac{dp}{dr} \right) \, ,
\label{Euler-Hydrostatic-Eq.}
\ee
which is the only non-trivial Euler hydrodynamic equation, as~seen in the matter-local reference frame (not by a distant observer) that relates the pressure gradient to the radial evolution of the metric component $\phi(t,r)$ characterising the gravitational collapse. This relation is understood as the ``equation of continuity'' for the fluid making up the QGP~core.

In addition, the~$G_{rr}, T_{rr}$ of Einstein's equations provides the other r-dependent $\frac{d\phi}{dr}$ metric field equation ($
\therefore e^{-\Psi} = 1 - \frac{2 Gm}{r c^2}$).

\begin{widetext}
\be
- \frac{8 \pi G}{c^4} p \,[{f\big(L(\tilde{\mu}(\hbar),B\big)} e^{\Psi}] = \frac{- {f\big(L(\tilde{\mu}(\hbar),B\big)} [r \frac{d\phi}{dr} + e^{\Psi}] - [{f\big(L(\tilde{\mu}(\hbar),B\big)}]^2}{r^2}.
\ee
\end{widetext}

Thus, by~resorting to the four equations from the diagonal $(tt), (rr), (\theta \theta), (\phi \phi)$ metric components in the Einstein Equation~(\ref{Einstein-Eqs}), plus the diagonal classical matter (M) and semi-classical (NLED) field terms in the action, i.e.,~the auxiliary link connecting the metric components with the fluid physical properties as of the continuity Equation~(\ref{T-conserv}) worked out above, one is left with five differential~equations. 

However, recalling that in general relativity the field equations are second-order differential equations, then only a couple of independent solutions are required. This leaves only three independent equations involving the $(tt), (rr)$ components of Einstein's Equation~(\ref{Einstein-Eqs}): $G_{tt} = \langle T_{tt} \rangle $, $G_{rr} = \langle T_{rr} \rangle $, in~addition to the continuity Equation~(\ref{T-conserv}), as~those remaining are independent of the azimuthal part of the gravitational field geometry (\ref{eff-metric}).

\begin{widetext}
\begingroup
\makeatletter\def\f@size{9}\check@mathfonts
\def\maketag@@@#1{\hbox{\m@th\normalsize\normalfont#1}}%
\be 
\label{modified-tov}
{\Big[  
{f(L(\tilde{\mu}(\hbar),B)} 
\Big]^{-1} }
\left(\frac{dp}{dr}\right) \left[ 1 - \Big[ 
{f(L(\tilde{\mu}(\hbar),B)}
\Big]^{-1} 
\left\{ { \left( \frac{ {L^2_{_P}}  }{2r} \right)} \frac{(\frac{dp}{dr})}{(\rho c^2 + p)} \right\}  \right] \, \,  =  \, \, - { G_{_{\rm N}} } \left( \frac{ m }{r^2} \right) \, \rho \left\{ \frac{ (1 + \frac{p}{\rho \, c^2}) \left(1 + 4\pi r^3\, \frac{p}{m c^2} \right)}{\left[1 - \frac{2 G_{_{\rm N}}  m}{ c^2\,r } \right] } \right\} \, .
\ee
\endgroup
\end{widetext}

Thence, by~considering the metric (\ref{eff-metric}), the~nonzero components of the Einstein \mbox{Equations~(\ref{Einstein-Eqs})} reduce to the first coupled Einstein equations. Finally, by~replacing the conservation Equation~(\ref{T-conserv}) into the components $G_{tt} = \langle T_{tt} \rangle $ and $G_{rr} = \langle T_{rr} \rangle $ of Einstein's field equations, we yield Equation~(\ref{modified-tov}). The~left-hand side in (\ref{modified-tov}) gathers the novel contribution to the stellar core dynamics from the vacuum polarization effects characterising the NLED. ($L_{_{\mathrm{P}}}$[m]=$\sqrt{\frac{\hbar G_{_{\rm N}}}{c^3}} \simeq 10^{-35}$\,[m]: Planck length. It absorbs the factor $\hbar$ from $\tilde{\mu}(\hbar)$.This is the nonlinear Tolman--Oppenheimer-Volkoff (N-TOV) equation~\cite{TOV} for studying figures of equilibrium of stellar fluids permeated by nonlinear electromagnetic~fields.

To completely determine the structure of a spherically symmetric body of isotropic material in hydrostatic equilibrium, the~TOV equation must be supplemented with an equation of state (EoS) relating the density of matter $\rho$ to pressure $p$, $F(\rho,p)\,=\,0$, e.g.,~a polytropic EoS: $p=K\rho^\gamma$, here $K$: polytropic constant, index $\gamma \equiv \frac{d(\ln{p})} {d(\ln{\rho})}$ and adiabatic index $\Gamma = \frac{\rho(p)\, c^2 + p}{\rho(p)\, c^2} \gamma$. In~our present case, both thermodynamic effects, vacuum polarization, and asymptotic freedom, are described by EoS: $p = - \rho (c^2)$, with~NLED magnetic pressure $p_{_{NL}}(B)=\frac{(B^{Star}_{QGP})^2}{2\mu_0}$. Consequently, a~non-polytropic EoS is demanded for the present astrophysical QGP star model. Thus, an~EoS such as the MIT Bag Model or some other kind can be called for.

Equation~(\ref{modified-tov}) is an algebraic quadratic equation for the pressure gradient ($\frac{dp} {dr}$) which renders a typical couple of solutions. Hence, the~second root of the N-TOV Equation~(\ref{modified-tov}) leads to the differential equation for the pressure gradient.
Furthermore, be aware that the entire term within the square root in  Equation~(\ref{ode-pressure-grad}) cannot be negative, otherwise $\sqrt{-1}$ = $i$ crops out. Thence, it should be equal to some constant $\Omega^2$, with~$\Omega\,>\,1$.

\begin{widetext}
\be
\left(\frac{dp}{dr} \right) =  \left\{ F\left(\frac{B^2}{b^2}\right) \right\}^2 
{\left[ \frac{r\,(\rho {c^2} + {p})}{ {L^2_{_P}} } \right] }
\left( 1 + {\underbrace{
\sqrt{1 + \left\{
\frac{ 2 \,G_{_{\rm N}} \, m  }{c^2\, {r}}
\right\} {\left(\frac{L^2_{_P}} {r^2} \right) }
{\left[ F\left(\frac{B^2}{b^2}\right) \right]^{-2}}
\left\{ 
\frac{\left[1 + \left(\frac{4\pi r^3\,p} {m c^2}\right) \right] } {\left(1 - \frac{2 G_{_{\rm N} } m} {c^2\,r}\right)} 
\right\} } }_{ \Omega^2 \,\,\, 
::: \,\,\, \Omega\,>\,1 } } \right) \, .
\label{ode-pressure-grad}
\ee
\end{widetext}

\hl{This equation is like Carballo-Rubio's (2018) analog derived within the study of the stellar structure in semi-classical gravity} \cite{Carballo-Rubio(2018)}, \hl{except for the NLED's effective metric contribution to the actual gravitational field around the QGP/QCD star in this model. This feature makes the present analysis a definitively original one.}

\subsection*{Solutions to Pressure-Gradient Equation: The Unsuspected Mass--Radius~Relation }
\label{sec:6}

From all this, closed solutions can be obtained for the thermodynamic properties of pressure and density, and~from Equation~(\ref{core-mass}) for the remnant core mass--the most relevant quantity searched for. 
Mathematically, the~method for obtaining solutions to the (NLED modified) pressure-gradient equation involves the rather standard Oppenheimer and Snyder TOV~\cite{Oppenheimer-Snyder(1939), TOV,  TOV-2}. 

The new solutions are obtained by drawing primarily on the constraint on $\Omega$ and the condition (i) stated above for the remnant surface pressure: $p$($r$ = $R$) = 0, in addition to~retaking the field equation for the ($tt$-component), so that the key expression for the GECKO's mass is given by Equation~(\ref{mass}).
The other thermodynamic properties $p(r,B^{Star}_{QGP})$, $\rho(r,B^{Star}_{QGP})$, as~well as the scalar function $\psi(r,B^{Star}_{QGP})$ and the metric function $\phi(r,B^{Star}_{QGP})$, can be easily~computed. 

\section{Discussion on the Mass--Radius~Relation }

Now, what is bluntly an unexpected result is what this Equation~(\ref{mass}) tells us about the structure (figure of equilibrium) of the terrifyingly hypermagnetized ultra-compact object we study here. The~terms inside the large parenthesis (\ldots ) add the first ``1'' to some wholly pure number; the part starting with the coefficient $\psi(r,B^{Star}_{QGP})$ and the long expression---inside the square brackets [\ldots ]---contains the contribution from the QGP star's extreme ultra-strong magnetic field ($B^{Star}_{QGP}$) for strengths well beyond the Schwinger limit~\cite{Broderick-etal(2000), Feynman-Electrodynamics}.

\begin{widetext}
\begingroup
\makeatletter\def\f@size{9}\check@mathfonts
\def\maketag@@@#1{\hbox{\m@th\normalsize\normalfont#1}}%
\be
\centering{  
M(r,B)\Big|^{Star}_{QGP} = \frac{c^2 \, r}{2G}  \left( \frac{ \left\{ 1 + \frac{ \psi(r,(B^{Star}_{QGP})) }  {(\Omega - 1)} \right\} }{1 - \left\{ (1 - \Omega^2) \frac{L^2_{_P}}{r^2} 
\left[1 - \left(2\underline{A} \frac{\mu_{_0}^{-1} (B^{Star}_{QGP})^2}{(b=B^{B-I}_{ ATLAS})^2}\right) + {\cal{O}} (\hbar^2) \ldots  \right]^2 \right\} } \right) 
}
\, \therefore {\rm for \,\, (10^{18} \leq B^{Star}_{QGP}(G) \leq 10^{22-42})} .
\label{mass}
\ee
\endgroup
\end{widetext}

which leads to the wide mass spectrum shown next and plotted in Figure~\ref{Fig.-4}; next,

\begin{widetext}
\begingroup
\makeatletter\def\f@size{9}\check@mathfonts
\def\maketag@@@#1{\hbox{\m@th\normalsize\normalfont#1}}%
{
\be
\label{mass-spectrum}
(0\lesssim M^{\rm{QGP}}_{\rm{Star}}\lesssim\,7\,M_\odot \longrightarrow), \,
\rm{stellar \,radii }  
\,(0\lesssim R^{\rm{QGP}}_{\rm{Star}}\lesssim 24\,km \, \longrightarrow), \,
field \, strengths \,  (10^{14} \leq B^{\rm{QGP}}_{\rm{Star}} \leq 10^{16}\,G \, \longrightarrow).
\ee
}
\endgroup
\end{widetext}

Poring over Equation~(\ref{mass}), it indeed defines the whole object's mass up to radius $r$=$R$ which, as~indicated above in discussing the solution to the TOV equation, defines the surface of the object; i.e.,~$m(r$=$R_{|_{\rm p=0}}$)=$M^{Star}_{QGP}$, where its pressure $p^{Star}_{QGP}$ vanishes. Conversely, the~first part of Equation~(\ref{mass}): 

$$\frac{2 G M(r,B^{Star}_{QGP})}{c^2} = r\cdot 1$$ 
is represented as though those terms were evaluated merely at the Schwarzschild radius $|_{\rm r=R_{Sch}}$, thus localising the surface that defines the standard true (theoretical) BH, i.e.,~the event horizon\footnote{It is a null surface characterized by null world-lines on it, all with ($\Delta \tau$ = 0) zero proper time, which is the amount of time elapsed along a world-line, i.e.,~what your wristwatch measures
\label{null-surface}} \cite{Gravitation(1973), Landau-Lifshitz(1979), landau-lifshitz(1970), Blau(2020), paper-ApSSc-2021, Ruffini-Wheeler(1971)}. 

Consequently, Equation~(\ref{mass}) decidedly pinpoints to an effective radius that is absolutely larger than the corresponding Schwarzschild radius for the same mass $M^{Star}_{QGP}$. A~result inasmuch as was categorically stated by Dirac in connection to singularities in space--times~\cite{Dirac-no-BH(1962)}: ``\ldots  So from the physical point of view, the~possibility of having a point singularity in the Einstein field is ruled out. Each particle must have a finite size no smaller than the Schwarzschild radius.'' Hence, Equation~(\ref{mass}) does not signal the formation of any frozen-future trapped (null) surface (see Note 3); the purported BH of mass $M^{Star}_{QGP}$, but~rather the appearance of a terrifyingly magnetized ultra-compact astrophysical object, a~QGP star, with~that mass and an extremely large, but~not infinite, surface gravitational redshift $z_{_{\rm Grav}} \gtrsim 10^8$. Recall that QGP (ALICE/LHC) is the unique experimentally probed physical state that nuclear matter inside an HHMNS can, in~principle, achieve in a stable manner. This characterizes its ultimate stage of stellar evolution over these extreme astrophysical conditions: a catastrophic hadron-to-quark deconfinement while pervaded by ultra-strong magnetic field strength that leaves it on the brink of an ultimate gravitational collapse, the~GECKO state. See Figure~\ref{Fig.-4} below and the arguments in its caption. (For a different conclusion suggesting BH formation in the process of material fallback onto a remnant ultra-massive core after the explosion of a 40\,M$_\odot$ supergiant star, see~\cite{chan-etal(2018)}).

Therefore, the~different take in this paper is that the astrophysical end state of the gravitational collapse must be a terrifically magnetized self-bound QGP BEC star, likely the Ginzburg's superstar~\cite{Ginzburg-1964, Thorne-1964, Thorne-1965}, but~not any true theoretical black hole, which, en passant, is defined as a non-stopping self-contracting space--time discontinuity (''an astrophysical object''?) \,\cite{Oppenheimer-Snyder(1939), TOV, TOV-2}. That is, the~astrophysical remnant should be a hypermassive QGP star supported in addition to gravity by the NLED powerful repulsive pressure acting on a QCD-dominated incompressible BEC of quarks and gluons held in such a state by the asymptotic freedom phenomenon: persistent scattering among parton-bound states, rather than producing a crack in space--time. That is what Einstein himself once called {\sl the Hadamard disaster} and to which he added: ``
\ldots General Relativity is about forces, not geometry.'' Additionally, last year in an arXiv paper, in~discussing the geometrical aspects and the physics/astrophysics that can take place in his own Kerr space--time regarding the dynamic role of frame dragging, Roy Kerr pointed out: ``\ldots  The~Kerr solution can be used to approximate the field outside a stationary, rotating body with mass $m$, angular momentum $ma$, and~radius larger than $2m$. \ldots then the rotational and Newtonian forces  outside the source drop off like $R^{-3}$ and $R^{-2}$, respectively. Clearly, spin is important close in, but~mass dominates further out \cite{Kerr-No-Singularities(2023)}.'' Certainly, a~rotating QGP star model must be what better fits in the actual dynamics we assess in this~paper.

Upon pondering all the above, in~our view, sound claims, one can ask whether Lifshitz and Khalatnikov were right when they stated~\cite{Lifshitz-Khalatnikov}: ``
\ldots  In~the universe, dust clouds or collapsing stars will expand again long before they reach the point of singularity\ldots '' From our understanding, they did have reason. Moreover, decades ago, Abramowicz, Kluzniak, and~Lasota claimed to have found no observational proof of identifying, and~an apparent impossibility to conclusively identify, the presence of BH horizons in BH candidates as far as electromagnetic radiation is used~\cite{Abramowicz-etal(2002)}. As~regards this last assertion, we recall here the EHT Coll. radio images of M87 and Sgr\,A$^\star$ \cite{EHT-BH-Radio-Images}, as~from them one cannot claim to have observed an actual astrophysical BH, but~rather only the presence of something equivalent to a photon sphere and an accretion~disk.

With that said, perhaps there is still room to check for echoes from the BH horizon via the detection of gravitational radiation by using advanced LIGO/VIRGO/KAGRA and other GW observatories so as to ``hear'' either some sort of echoes from the abyss (see Afshordi~et~al.~\cite{afshordi-etal(2021), afshordi-etal(2021)-1, afshordi-etal(2021)-2, afshordi-etal(2021)-3}, and~Cardoso~et~al.~\cite{cardoso-etal-2016}, and~Olivares~et~al.~\cite{Olivares-etal.(2020)}) or a GW reverberation-like effect. Unfortunately, hitherto non-positive results were pointed out by the very recent study case in~\cite{KAGRA-Results(2023)} and references therein, which indicates that ''\ldots no significant echo signals are found from O3 LIGO-VIRGO-KAGRA Coll. detected GW~events\ldots ''.

This interplay of ultra-relativistic gravity and the extreme repulsive pressure of the yo-yo state emulating a dark energy-dominated cosmological fluid ($p=-\rho$) \cite{A-P.Phys(2011), hjmc-gaetano(JCAP2011), hjmc-gaetano(PRD2009)} set up an oscillation that creates the equivalent of ultra-high-frequency sound waves in the QGP star BEC state. (Quite unlike the process that drove Baryon Acoustic Oscillations (BAOs) in the very early universe, producing expanding bubbles~\cite{peacock(1998)}). In such a unique nuclear astrophysics environment---a yo-yo state super-powered by gravitational attraction---these waves forcefully become a quasi-resonant state that increases  the already gigantic repulsion pressure even more to stably sustain the compact hypermassive QGP~star. 

To ascertain the details of the crucial role played by the extremely high magnetic field $B^{Star}_{QGP}$, keep in mind how essential it is for both the emergence of the vacuum polarization state and its related huge repulsive pressure, which are able to refrain the definitive collapse to the theorists' BH, and~also for the actual impressive amplification of its field strengths up to that represented by the unitless NLED-derived factor 
$F \left(\frac{(10^{18})^2 \leq (B^{Star}_{QGP})^2 \leq (10^{22})^2}{(b=B^{B-I}_{ATLAS})^2} \right)$---part of the $\psi(r,B^{Star}_{QGP})$ function in Equation~(\ref{mass})---in view of the BEC of quarks and gluons appearing when the subgroup Z$_3$ symmetry gets broken inside the HHMNS (see Figure~\ref{Figure 1}). This is a typical instance of self-magnetization from charged quarks subsumed by Bose--Einstein condensation. That is, when the HHMNS magnetic field grows up to the limit indicated above ($B^{Star}_{QGP}$), it enhances by powers of ten the term $\psi(r,B^{Star}_{QGP})$  in Equation~(\ref{mass}). This makes the radius of the resultant QGP star ``significantly'' larger, as~compared to the factor associated with the Planck length $L_{_P}$ alone.

\begin{figure*}
\includegraphics[width=12.0cm]{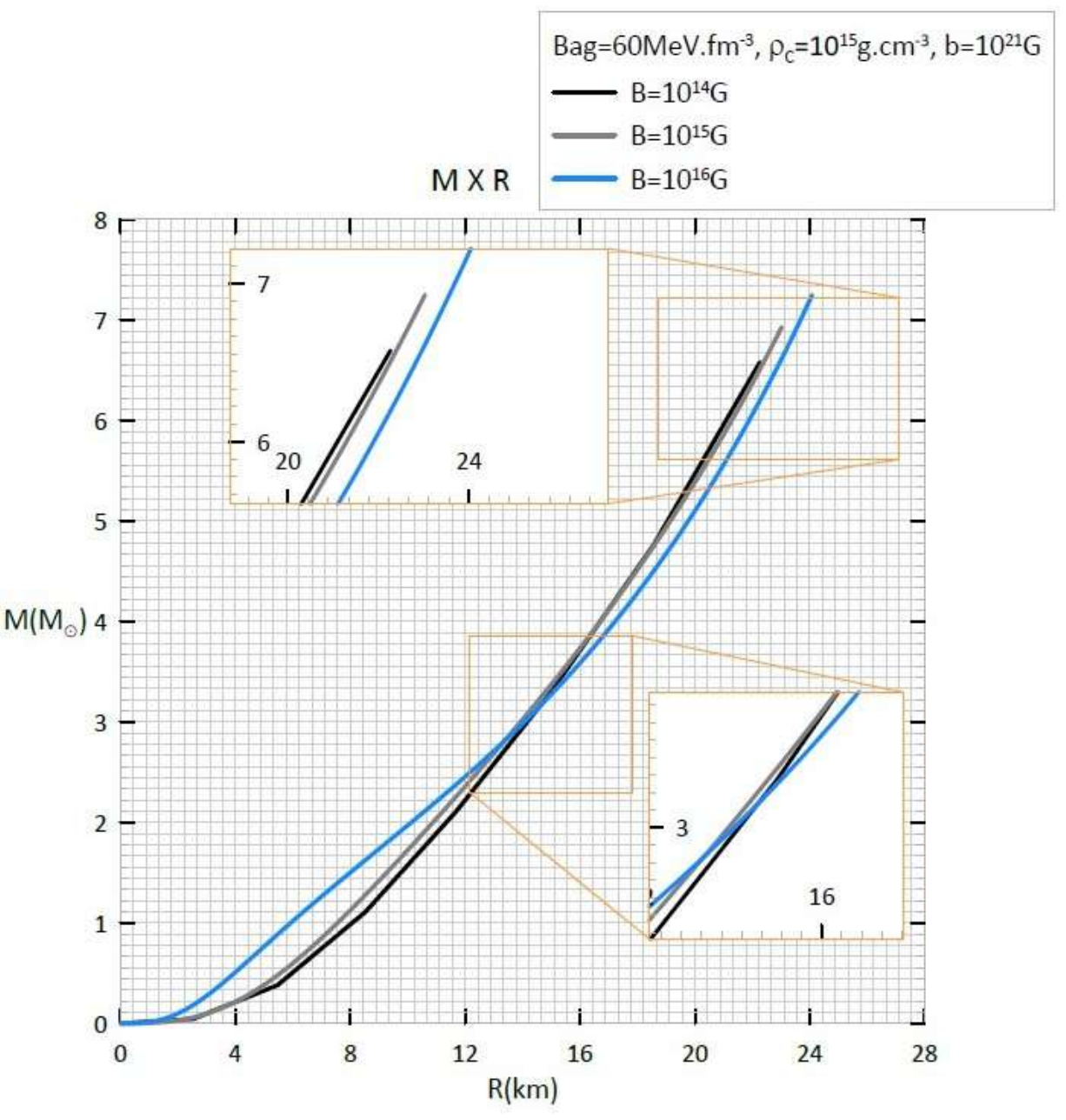}
\caption{A sample of the mass vs. radius relations  calculated for the QGP Star model according to the physical framework
here presented. For~this, the~ MIT Bag Model EoS with specific values for the central density $\rho_c$ and the B$_{bag}$ constant was used. The~relations are obtained by satisfying the conditions that define the actual stellar radius ($\rho(r)|_{r=R^{QGP}_{Star}}, p(r)|_{r=R^{QGP}_{Star}} = 0$!) as a function of the magnetic field strength $B$ at the QGP star surface, and~for a field strength  maximum value $b$ in the Born-Infeld theory of nonlinear electrodynamics (NLED). This graph exhibits a fundamental trend indicating that there appears not to exist 
any mass/radius upper bound to the general relativistic hypermassive (up to $7\,M_\odot$ in this plot) QGP star
figure of equilibrium.
This also means that the quark-gluon plasma can built astrophysical objects with masses spread over a wide spectrum so as to include the central supermassive compact bodies observed in most galaxies.
A wide range of models exhibiting the relentless trend pictured in this graph will be discussed in forthcoming papers wherein the parameter space involving the EoS, the~field strength at both QGP star core and surface, as~well as the NLED Born-Infeld theory maximum field strength $b$ (including the upper threshold derived from the ATLAS/LHC experiment, as~quoted above) are explored. [This plot  is taken from the paper in preparation by H.J. Mosquera Cuesta, R. Francisco dos Santos and L.G. de Almeida: General relativistic figures of equilibrium of anisotropic hypermagnetized quark-gluon plasma stars as the ultimate astrophysical state of gravitational collapse supported by NLED-driven vacuum polarization. In~that paper, a wide number of graphs describing key relations (e.g., density vs. radius, pressure vs. radius, compacticity (Buchdahl limit), etc.) among the QGP star astrophysical properties are given, along with an extended analysis of the physical implications of the results obtained within this scenario. }
\label{Fig.-4}
\end{figure*}


\section{Summary}

Much as the ATLAS/LHC (2018) results allowed us to nail down the reality of the nonlinear phenomenon of photon--photon scattering as NLED foresaw it, similarly, \mbox{NLED/QED + QCD} also appears to dominate the final stage of the general relativistic gravitational collapse. 

It startlingly states that the radius of the generic figure of equilibrium of ultra-compact hypermassive extremely magnetized stellar quark cores is always larger than the Schwarzschild radius for any given mass. 
{Indeed, the~calculated spectrum covers a wide range of masses, radii, and~field strengths:}


\begin{widetext}
\begingroup
\makeatletter\def\f@size{9}\check@mathfonts
\def\maketag@@@#1{\hbox{\m@th\normalsize\normalfont#1}}%
{
\be
\label{mass-spectrum}
(0\lesssim M^{\rm{QGP}}_{\rm{Star}}\lesssim\,7\,M_\odot \longrightarrow), \,
\rm{stellar \,radii }  
\,(0\lesssim R^{\rm{QGP}}_{\rm{Star}}\lesssim 24\,km \, \longrightarrow), \,
field \, strengths \,  (10^{14} \leq B^{\rm{QGP}}_{\rm{Star}} \leq 10^{16}\,G \, \longrightarrow)\, ,
\ee
}
\endgroup
\end{widetext}
{
which, as~is clearly seen in Figure \ref{Fig.-4}, exhibits a relentless trend to higher masses and their corresponding radii with apparently no bounds to those physical properties. } 

This finding is along the lines of the recent analysis in Ref.~\cite{Kerr-No-Singularities(2023), Mazur-Mottola-GRAV_BEC_Object(2023), Lifshitz-Khalatnikov, Abramowicz-etal(2002), Dirac-no-BH(1962), Carballo-Rubio(2018), cardoso-etal-2016, Olivares-etal.(2020), afshordi-etal(2021)}. 
Notwithstanding, notice that ours' goes well beyond from the perspective of of ``materialising'' the quantum nature of the collapsed ultra-compact astrophysical remnant, {the super strongly magnetized QGP star. 
It also highlights the fundamental concomitant roles played by both the vacuum polarization driven by NLED and the  asymptotic freedom of quantum chromodynamics in precluding the definitive collapse~\cite{Lifshitz-Khalatnikov} to the purported singularity.
 }

In our view, the~astrophysical dynamics described above are quite close to revealing the actual mechanism responsible for the endless compression admission caused by the injection of gravitational energy to the steadily density-changing QGP (BEC) star via the gluon-mediated enduring exchange of color charge among bound states, e.g.,~the odderon: a 3-gluon glueball state. 
{
The last is a unique particle formed by multi-interaction among gauge bosons that causes differences in proton--proton and proton--antiproton scattering  (TOTEM/LHC/CERN and $D\emptyset$/Tevatron/FERMILAB Collaborations) \cite{the-odderon-discovery, antchev-etal(2020), Csorgo-etal.(2021)}.}~{Another possibility is the formation of other coalescence states like the lightest pseudo-scalar glueball X(2370) that was discovered at BESIII (2011) and had its nature recently confirmed \mbox{(2024)  \cite{PRL-132-181901-(2024)}. }}

Incidentally, all of these physical insights place a great burden on the theorists' idea of the trapped null surface, i.e.,~the true theoretical black hole, which on the basis of the present analysis cannot turn out to be a proper astrophysical body the way relativity theorists idealistically deemed it to be decades ago. Our new analysis suggests otherwise. Indeed, Roy Kerr stated in his paper~\cite{Kerr-No-Singularities(2023)} that: ``\ldots  There is no known reason why there cannot be a fast rotating non-singular star inside the horizons generating the Kerr metric outside. There is no published paper that even claims to prove that this is impossible and yet so many believe All black holes contain a singularity.''  
The present theoretical study, en passant, confirms, from~a very different physical approach,  the~similar conclusions achieved by other authors~\cite{Mazur-Mottola-GRAV_BEC_Object(2023), MPL-A(2010), Carballo-Rubio(2018), cardoso-etal-2016, Olivares-etal.(2020), afshordi-etal(2021)} and the recent article in Ref.~\cite{paper-ApSSc-2021}. Perhaps a new kind of physics/relativistic astrophysics is~afoot.

\label{sec:A}
\section{More on Magnetic Field Amplification Driven by Gravitational Collapse}

Despite the fact that the inner magnetic fields of neutron stars have not yet been quantified through direct observations\footnote{It still appears to be room for estimating, via direct astronomical observations, the~B-field strength of an ultra-magnetized QGP star by extracting the total chemical potential and, in~particular, the~contribution to it from the magnetic field present during the phase transition to quark--gluon plasma. See~\cite{measuring-chem-pot-by-astron} which introduced such a method and our forthcoming paper on this promising avenue.}, resorting to the scalar virial theorem, flux conservation, allows us to estimate that such fields can reach values up to $B_c \simeq 10^{18}$\,G in the centre of stars. Such extremely strong magnetic fields, which can be characterized as stochastic due to the natural stochastic fluctuations in the stellar electron--ion plasma, represent the time-dependent radial magnetic field imposed by the dynamics of the collapsing star we focus on hereafter, regardless of whether it is actually monopole, oblique dipole, or~multipole~\cite{Petri's-papers(2012-2016), Petri's-papers(2012-2016)-2, Petri's-papers(2012-2016)-3}. Notice that for an extremely relativistic ultracompact object (R\,$\lesssim R{\rm{^{Typ}_{NS}}}$), the~polar magnetic field B$_p$ becomes much weaker than the equatorial field B$_e$, i.e.,~B$_p$/B$_e\sim\,10^{-8}$, all in direct contrast to their corresponding Newtonian behavior B$_p$/B$_e\sim\,2$ (see~\cite{paper-ApSSc-2021} and references therein). This extreme asymmetry in field configurations induces a marked anisotropy between axial/polar vs. radial/equatorial pressures~\cite{IJMPD(2008)}. This physical effect was explored decades ago in~\cite{Quant-magn-gravit-collapse, EPJ-C(2003)}
and since then by a large number of researchers as a mechanism to drive a catastrophic implosion, kind of a quantum magnetically induced collapse~\cite{IJMPD(2005)} in hypermagnetized neutron stars/pulsars to render a quark star as its astrophysical remnant. This summary is enough for now; we shall come back to this essential issue~later. 

Aside from this, the~extreme asymmetry in the magnetic field structure can also be regarded as an astrophysical tool to pin down the GECKO state (gravitationally eternally collapsing ''kompact'' object) on the brink of definitive collapse by, for~instance, resorting to either spectroscopic line analyses or searching for exotic synchrotron radiation spectra, as~well as extreme ultra-high energy cosmic ray (including neutrinos) emissions from prospective unknown~sources.

With that in mind, in~this scenario, the GECKO survives in a suspended-collapse state. Such a state is similar, in~nature, to~the continued gravitational contraction studied by Oppenheimer and Snyder in 1939 \cite{Oppenheimer-Snyder(1939), TOV,  TOV-2}, but, in~contrast to such non-stopping dwindling (built on a nonphysical assumption of pressureless matter---dust)\footnote{The unavoidable aftermath of such an advantageous---dust---approach involves reaching an infinite energy density at the centre of the star, a~sort of a singularity. However, as~discussed by Faraoni and Vachon \cite{V-Faraoni(2020)}, dust particles follow geodesics, but~for a general perfect fluid endowed with pressure (as in our model here), the~proper time along the trajectory does not coincide with the proper time along the corresponding geodesics. Notwithstanding, a~4-force parallel to the trajectory of a massive particle can always be eliminated by going to an affine parameterization, but~the proper affine parameter is always different from the proper time.} the GECKO is blocked out altogether, remaining in an endless shrinking vs. expanding yo-yo state---a configuration prompt to generate GW in the NLED-induced spacial asymmetry---by virtue of the strength of gravity and the concomitant action of both the NLED vacuum polarization ---triggered by such extremely ultra-strong magnetic fields---and the QCD repulsive potential. This is a critical example of self-magnetization by quark condensation that produces repulsive scattering pressure via the characteristic positive potential energy of the QCD asymptotic freedom. But~more crucially, it also brings in a repulsive force among quarks and gluons, impeding the infinite self-contraction (see Figure \ref{Figure 1}). Both effects produce a total negative pressure of the type $p$\,=\,$-\,\rho$, reminiscent of the Chaplygin-like gas\,\cite{Chaplygin-Gas_Moschella-etal.(2003)}, akin to Chandrasekhar's quantum degeneracy pressure---a result of Pauli's Exclusion Principle---that opposes a definitive implosion or ultimate gravitational collapse to the closed-future trapped (null) surface, the~theorist's true black hole~\cite{Ruffini-Wheeler(1971), Gravitation(1973), Blau(2020)}. (Note: The black hole concept was introduced by J. A. Wheeler in an international conference on general relativity  held in Chicago in 1967 and then by R. Ruffini and J. A. Wheeler in a  1971 paper~\cite{Ruffini-Wheeler(1971)}). The physical effects of an near-repulsive-pressure BEC were also explored by Mazur and Mottola in their spherically symmetric de-Sitterlike inner gravitational vacuum condensate---gravastar---as an alternative to BH (see their recent review~\cite{Mazur-Mottola-GRAV_BEC_Object(2023)} and references therein), which also satisfies the EoS: $p_{_V}$\,=\,-$\rho_{_V}$ $\longrightarrow V(\phi) > 0$. (Check the next footnote about potentials in physics)\footnote{It is worth remembering that in physics, most potentials are negative and attractive---defined as ${\rm V(r)<0 \therefore V(r)=-{\cal F}(r)}$, e.g.,~the electrostatic (Coulomb) potential, the~gravitational potential, or~the elastic potential describing a deformable mechanical spring (Hooke's law). Indeed, signs set out a physical fundamental difference. For~a vector theory---like electromagnetism---the potential energy of the field among similar charges is positive and increases as their inter-distance decreases; i.e.,~one needs to invest work to push similar charges together; as a result, the~force among them is repulsive. Conversely, for~a tensor theory---like gravitation---where the charges are the masses themselves, the~field potential-energy term is negative, which means work has to be injected in order to enlarge their separation. That is, the~force among similar charges is attractive. The~same behaviour holds for unlike charges in electromagnetism. Nonetheless, exceptions do occur in physics, and~some natural phenomena are pictured differently. There exists a noticeable potential that combines both the repulsive ($1/r^n$) and the attractive ($-\,1/r^m$) properties laid out above, the~Lennard-Jones potential: a simplified, radially symmetric model describing the essential features of interactions between neutral pairs of either atoms or molecules: $V(r) = 4\,\epsilon\, [(\sigma/r)^{12} - (\sigma/r)^6]$ $\therefore \sigma$: distance at which the particle-particle potential energy $V = 0$, and~$\epsilon$: depth of the potential well, which has its minimum at: $ r=r_{\rm {min}}=(2^{\frac 1 6})\,\sigma$, where $V = - \epsilon$. Two interacting particles repel each other at very close distances, attract each other at moderate distances, and~eventually stop interacting at infinite distance.}. Thus, in~our analysis, the~GECKO's effective mass ends up being increased by either the Landau quantization~\cite{Ginzburg-1964} or the vacuum awakening via quantum fluctuation~\cite{matsas-etal.(2010), matsas-etal.(2010)-2} mechanisms, and~consequently 
might in fact end up as the Ginzburg's ultra-magnetized superstar, an~object which construes Thorne's argument that according to ``\ldots  the principle of flux resistance to gravitational collapse\ldots  in the presence of particulate matter\ldots  no pure magnetic energy can collapse into a BH state.'' Indeed, Thorne stressed that his analysis does not apply when the fields reach the critical 
Schwinger value and its associated polarization effects~\cite{Thorne-1965}.

In general settings, at~the innermost core, as~those field strengths go beyond the Schwinger limit, it happens that the quantum vacuum---described by QED as a current/charge-free, polarized, and~magnetized medium---undergoes a sort of ``awakening'' through the vacuum polarization phenomenon. The~awakening of the vacuum also increases the core’s effective mass \,\cite{Ginzburg-1964, matsas-etal.(2010)}. In the scenario portrayed by the present paper, this is a critical contribution of extreme relevance to the actual HHMNS mass at the end state of gravitational collapse and catastrophic color deconfinement. The question reads: In an exotic fluid such as this, dominated by a total negative pressure $p = - \rho$, what amount of gravitational mass can be stably afforded to form a QGP star, as~we model it here? Perhaps the scalar Virial Theorem could guide us in this critical direction if one properly includes the quantum vacuum fluctuations as suggested in~\cite{matsas-etal.(2010)}, where there appears not to exist a limit to the energy density exponential growth of such a squeezed quantum vacuum driven by the super strong---twisting---gravitational field of the collapsing matter. (Notice that an uncontrolled energy outburst may destruct the compact star expected to form.) At such a stage, a~gravitational mass bound on the stable hypermassive QGP star---as idealized in this paper---might appear by releasing the over-increase in energy density 
--due to the ``awakening''---via a massive/powerful generation of gravitational radiation (recall that NLED breaks down the Schwarzschild spherical symmetry) in the yo-yo state figured out above, or~by radiating the hypothetical dark photons, or~from achieving the energy scale theorized to start to radiate axions---a Goldstone particle resulting from a broken U(1) global symmetry in the Peccei--Quinn mechanism to cancel the anomalous Charge Parity (CP)-violating term in the QCD Lagrangian, making QCD CP-conserving~\cite{Axions-Conference(2024)}---or, in an extreme case, from~the higher threshold imposed by the GUT gauge symmetry-breaking energy scale. In~this astrophysical environment, the~hypercritical magnetic field dumps its gigantic energy reservoir into radiation and matter by ``materialising'' itself via the creation of $e^-,e^+$ pairs and photons that take over due to the relatively huge strength of electromagnetic interaction with respect to both gravity and electroweak interactions. (Bear in mind that neutrino pairs $\nu^+,\nu^-$ are less likely due to the rules of symmetry: both the feebleness of electroweak interaction and their nonzero rest mass, in addition to~the stupendously low thermodynamic temperature of the astrophysical GECKO. All of this makes it a little harder to ever produce them.)\footnote{\label{Specific-Heat}
At this point, it is necessary to point out a key thermodynamic property: If a self-gravitating system loses energy, for~instance, by~radiating energy into space, the~case of star-like objects, then it contracts itself, but, in~contrast to standard substances, its average kinetic energy indeed increases (i.e., $\Delta Q_{Syst}<0 \rightarrow \Delta E_{Kin}>0 \rightarrow \Delta T_{Syst}>0$). As~the temperature is defined by the average heat content in the kinetic energy, the system can be said to possess negative specific heat capacity. The~more extreme version of this feature happens with excessively ultra-compact astrophysical systems, such as black hole-like objects, including the self-bound QGP star in its BEC-like state. According to black hole thermodynamics~\cite{S.Carlip(2014)}, the~more mass/gravitational energy a black hole absorbs, the~colder it becomes. Conversely, if~it is a pure emitter of energy, for~example through Hawking radiation, as~shown in a 1975 paper~\cite{S.W.Hawking(1975)}, it will become hotter and hotter until it finally boils away.} 

\section{Color Deconfinement Transition to QGP Star: Role of the QCD Asymptotic~Freedom}
\label{sec:3}

Quantum chromodynamics (QCD: see the pertinent review by Wilczek (2000) \cite{Wilczek(2000)}) describes both the strong force---as a classical approximation: F$_{\rm {St}} = - \nabla\,{\rm V}(\phi) = \alpha_{\rm {St}}\,\hbar\,c\, (4/3) / r^2$, with~$\alpha_{\rm {St}} >> \alpha_{\rm {em}}$---and the unique positive binding energy ${\rm V}(\phi)>0$ of quarks that attract each other to build baryonic nucleons---protons, neutrons---and their residual attraction inside atomic nuclei (see Figure \ref{Figure 1}). 

The QCD fundamental hallmarks~are: 
\begin{itemize}
\item[(a)] 
Asymptotic 
freedom, which refers to the vanishing of the strong nuclear (force) interaction among quarks as their inter-distance dwindles but never actually reaching zero (see Gross and Wilczek~\cite{Gross-Wilczek(1973), Gross-Wilczek(1974)}, and~Politzer~\cite{Politzer(1973)}, and~Figure \ref{Figure 1}).
\end{itemize}

\begin{itemize}
\item[(b)] 
Chiral symmetry breaking, the~dynamic spontaneous breaking\footnote{The opposite of explicit symmetry breaking of a theory happening in adding terms to its defining equations of motion, i.e.,~Lagrangian/Hamiltonian, such that the new equations do not preserve the posed symmetry.} of the symmetry appearing on the limit at which the quark masses are set equal to zero. It explains how quarks generate the masses of hadrons (see the review by Hooft~\cite{t'Hoof(1980)}).
\end{itemize}

The terrifyingly high inner-field strengths $B^{\rm inner}_{\rm core}\sim10^{18-20}$\,G and the overwhelming gravity pressure of supermassive ultra-compact cores\footnote{Keynote: as the gravitational field reaches the state when its strength is comparable to other forces, its putative quantum nature can no longer be ignored. That is why NLED is called for in this paper, as~the first step to integrating quantum effects into the whole astrophysical picture, as~we do not yet have a quantum theory of gravitation to hand.} also contribute to destabilising their innermost central region by compacting the HHMNS' nucleonic matter, causing the hadronic gas to undergo a catastrophic rearrangement---color-charge deconfinement transition---to a new phase, freeing their quarks and gluons---collectively known as partons~\cite{Asymptotic-freedom-in-parton-language, altarelli-parisi(1977)}---to constitute a ''hot'' (${\sim}10^{12}\,$K) dense medium, a~state that emulates the ALICE's QGP featured earlier. Surely the strange quarks, and~likely the charm quarks, are everywhere, as~they pre-existed in the transforming HHMNS. For~instance, in~(anti ${ \left(\frac{3}{\Lambda}\bar{H}\right)}$) hypertriton ${ \left(\frac{3}{\Lambda}{H}\right)}$ states---unstable nuclei composed of (anti)protons, (anti)neutrons, and  (anti)Lambda---have greater binding energy in strange nucleons---Lambda baryonic hyperons, as~discovered by the STAR Collaboration with the Relativistic Heavy-Ion Collider (RHIC) at Brookhaven National Laboratory (BNL) \cite{STAR-RHIC(2010)}. In~the case of an ultra-compact star core, this stage can be described by an EoS built up on the relativistic mean-field (RMF) model plus an extended MIT bag model~\cite{Greiner-Catastrophic-Phase-Trans,Greiner-Catastrophic-Phase-Trans-2, horst-stoecker-etal, Aurora-2021}, or~similarly via a symmetry-restoring chiral mean field (CMF) nonlinear sigma model \cite{Veronica-PhD-Thesis, QCD_Probes-AdS_QCD(2024), Blaschke-etal(2019)}\footnote{\label{color-deconfinement} First-order phase transitions are those that involve a latent heat. In~a compact system, the~matter gets heated up as gravity pulls it together (its negative specific heat at work). During~such a transition, a~system either absorbs or releases a fixed (typically large) amount of free energy per volume. In~systems transitioning to deconfinement, the~phases are connected at the chemical potential in which the pressure from the quark EoS matches/exceeds the one from hadronic EoS, pinpointing the phase transition to the quark--gluon plasma (QGP) state~\cite{QCD_Probes-AdS_QCD(2024), Veronica-PhD-Thesis}. Such an extension of hadrons-to-quarks is similar to the Nambu--Jona-Lasinio (NJL) model extended up to include the Polyakov loop ($\Phi$=0: $\Phi$ relates to free energy), that is, a~nonlinear sigma model that uses the Polyakov loop as the order parameter to feature the deconfinement (the ALICE results show up at the energies $\gtrsim$\,150\,MeV---a rightward shift of the Lattice QCD borderline in Figure 
 \ref{Figure 1}): $\scriptscriptstyle{\left(\Phi \,=\, \frac{1}{N_c} {\rm Tr}\, \left[{\rm P\, exp} \left(i \int_0^\beta d\tau {A_4}_{|_{iA_0}} \right) \right] \,=\, e^{-F_q} 
\Big\{_{\, 1 : \, F_q \rightarrow 0 \,\,\, \rm deconfinement}^{\, 0 : \, F_q \rightarrow \infty \,\,\, \rm confinement}\right).}$ \linebreak This is quite a natural next step, as~the Polyakov loop is related to the Z$_3$ symmetry (see Figure \ref{Figure 3} the vortex percolation/de-percolation threshold, as~described in QCD-Lattice gauge theory), which is spontaneously broken by the appearance of the quark condensate phase, which then makes ($\Phi$\,$\neq$\,0) finite. (See further details in Ver\^onica Dexheimer's (2009) PhD thesis~\cite{QCD_Probes-AdS_QCD(2024), Veronica-PhD-Thesis}). Such work, though, neither included Einstein's gravity nor even Maxwell's electrodynamics. Nonetheless, both physical key ingredients were included in a later paper. See \cite{papers-NS-deformation-1}). In~such a state, gluons also play a key role in both the entropy and baryon density via the Polyakov loop and its potential (V$(\phi)$, which imitates the effects from confinement; see Figures \ref{Figure 3}  and \ref{Figure 1}, with~$\phi$: the background color-charge gauge field with which quarks interact) and represent color-bound states that mimic extra possible states such as the QCD higher resonances, which last for  $\Delta$t\,$\lesssim$\,$10^{-24}$\,s (Delta baryons $\Delta^{++, +, 0, -}$, upsilons (mesons of the bottom quark $\Upsilon$($b\,\bar{b}$)) pinpointing to the phase transition to a quark--gluon plasma state, or~the spin-1 (vector) charged rho mesons (isospin-$\pm1,0$ triplet of states $(\rho^+,\rho^0,\rho^-)$ with $\Delta$t\,$= 4.41\times10^{-24}$\,s, a~result of chiral symmetry breaking). But~notice that mesons with a top ($t$) quark are very likely impossible, as~high -ass $\rightarrow$ rapidly decays before it has time to form: $\Delta$t~$\sim$\,5$\times$\,$10^{-25}$\,s.)
The Polyakov potential was built for analysing Lattice QCD at nil/low chemical potential and high temperature but was redesigned to discuss neutron star (NS) dynamics where high chemical potential and stupendously low temperatures are dominant, as~explained in Footnote (7). 
The Polyakov loop has key roles in the NS context: (a) for an increasing temperature/density, it takes on nonzero values. (b) It appears at the high-value coupling constants of both baryons ($g_{b\Phi}\,\Phi^2$) and quarks ($g_{q\Phi}\,(1-\Phi)$). Its presence in the baryons' effective mass hints at the suppression of baryons at the quoted threshold.  Meanwhile, its appearance in the effective mass of the quarks ensures that no quarks will show up at low temperatures/densities~\cite{QCD_Probes-AdS_QCD(2024), Veronica-PhD-Thesis}. 
In the present paper, we offer general insight into this unprecedentedly known fundamental result using our semi-classical approach. We postpone forthcoming research involving a full detailed analysis of this pathway to the astrophysical end state that follows the catastrophic deconfinement transition from the HHMNS to the QGP star, which we plan to model via either a Polyakov-loop-inspired NJL-like RMF model or the symmetry-restoring CMF Lagrangian, like the following one: 

{\scriptsize{
$$L = \frac{R}{2\kappa} + L{(F,\tilde{G})} + L^{\rm H,Q,L}_{\rm Kin} + L^{\rm B,(Q),V,S}_{\rm Int} + L^{\rm Self}_{\rm Scal} + L^{\rm Self}_{\rm Vec} + L_{\rm SB} -  V(\phi)\, .$$}}

\noindent The terms here represent Einstein's gravity, NLED, the kinetic energy of hadrons (H); quarks (Q) and leptons (L); interactions between baryons (B and quarks) and vector (V) and scalar (S) mesons; self-interactions of scalar (S) and vector (V) mesons; the term explicitly breaking/restoring chiral symmetry (SB); and the Polyakov-loop potential (V($\phi$)).
Notice, en passant, that a study of neutron star equilibrium configurations joining the four interactions was performed by Prof. Ruffini's group in \cite{belvedere-etal-nuc-phys-a}, the very first of its kind. Nonetheless, they did not take into consideration NLED, neither the hadron-to-quark deconfinement nor a high-mass HHMNS model. 
As befits a model of this sort, its stability can be explored based on causality conditions, the~adiabatic index, the~generalized ($t,r$-dependent) Tolman--Oppenheimer--Volkov (G-TOV) equation, Herrera's cracking method, or~the Buchdahl limit.}.

The shrinking core becomes even tighter (see Figures \ref{Figure 3} and  \ref{Figure 1}) but is still able to support itself due to both states: the quantum vacuum repulsive (negative) pressure and the asymptotic freedom property of parton matter, whilst the QGP star's high energy density reaches many orders of magnitude beyond that for atomic nuclei and typical neutron stars\footnote{In general settings (i.e., non-spherical/anisotropic), the~static configuration is achieved asymptotically and settles to a final radius greater than the photon sphere for the Schwarzschild black hole. The~central density becomes arbitrarily large and approaches a sort of naked singularity~\cite{Joshi-Malafarina(2011), joshi-book(2007), joshi-narayan(2011)}.}.~Specifically, models where the quarks interact strongly predict a harder-to-compress form of matter, a~feature that correlates with HHMNS of larger radii~\cite{ATLAS/LHC-Coll.(2017)}. For instance, a~massive NS of radius 12\,km and mass 2\,M$_\odot$ has a quark core $\sim 6.5$\,km. This is a model built with a subconformal EoS with a speed of sound of $c^2_{s}<(1/3)\,c^2$ \cite{Annala-etal.(2020), annala-vuorinen(2023), Aurora-2021}.

However, the~astrophysical state can worsen even further with regard to the standard picture of HHMNSs. In~the case of persistent mass accretion onto the hypermassive QGP star (${\rm 2\,\leq M^{Star}_{QGP}(|{M_\odot}) \simeq 5}$), such a suspended state of ultimate collapse is not everlasting. An exceptional role, perhaps, is reserved for either axions~\cite{Raffelt-etal-2008, raffelt book(1996)} or the unique nonzero-mass hidden/dark photons~\cite{dark-photon-mass}, if they do occur in nature, as~these extra losses of energy would set off unknown/unexpected dynamical effects. (This remains unresolved) In some instances, quantum-gravitational effects (QCD asymptotic freedom + NLED/QED vacuum polarization + relativistic gravity) can take control and exact a halt, potentially reversing it~\cite{Lifshitz-Khalatnikov} in a dismantling nonpareil explosion that may rival the universe's latent energy released at the phase of the spontaneous breaking of GUT gauge symmetry: $E_{\rm GUT}\sim 10^{19}$\,GeV, the~next energy scale higher than the one for the electromagnetic interaction, according to the standard model of particle physics. Though not yet available at particle colliders, and~perhaps never reachable apart from ~at cosmic scales, there is certainly room for it to be achieved in astrophysical or cosmological processes. Such unique explosions can be unequivocally disentangled among similar gigantic cosmic outbursts. Thus, perhaps some of the fourteen bright celestial sources of gamma rays (observed by the FERMI GRB Space Observatory) claimed to have come from stars made of antimatter~\cite{antimatter-stars} may instead come from the formation of the QGP star, i.e.,~the GECKO. Is there any signature of something having been left from such antimatter stars? Up to today, nothing has been identified that is compatible with the idea of antimatter stars. Neither the putative (theoretical) black hole, as~described in~\cite{Gravitation(1973), Blau(2020)}, nor any other remnants. En passant, within~the framework of NLED, some discussions on fundamental issues related to theoretical BH physics and particle astrophysics such as irreducible mass, neutrino propagation inside extremely dense star cores, etc. are given in~\cite{NLED-MORE-PAPERS, hjmc-etal(2017), hjmc(2017)}).


\section{Minimal Coupling of Gravitation to NLED: The  Lagrangian Theory, the~Effective Metric, and Einstein's Field~Equations}
\label{sec:4}
\unskip

\subsection{Born--Infeld Lagrangian for Featuring the QGP Star Extreme Electromagnetic~Fields}
\label{subsec:1}
Born--Infeld (B-I) nonlinear electrodynamics, a~field theory exhibiting relativistic structure and describing no birefringence \cite{Born-Infeld(1934)}, was addressed, just after the release of ALICE results by~Ellis, Mavromatos, and~You in connection to setting a strong bound on the theory's mass scale: ${\rm m^{B-I}_{ATLAS}}\sim$\,90\,GeV\,\cite{Ellis-etal.(2017)}, which, by using a conversion factor \cite{Jaffe(2007), wilczek(2007)} from natural units to magnetic field $B$ units---Gauss (G)---defines the field strength's \linebreak empirical limit:

\be 
b = {\rm B^{B-I}_{ATLAS} \simeq }\,(3.28) \cdot 10^{22}\,{\rm G} \, \left(\frac{\rm {\rm m^{2\,B-I}_{\,\,ATLAS}} = (90\,GeV)^2}{\rm GeV^2}\right)\,.
\ee
{It is worth noting that the Born-Infeld Lagrangian was built in part with Einstein's principle of relativity in mind: no material body can travel faster than light in a vacuum\footnote{There is no Lorentz transformation to turn a slow speed motion into one of a speed faster than light.}; and the ``on the mass shell'' $p_\mu p^\mu = m^2$ constraint, which also exhibits the self-duality~property.}

The mathematical structure of B-I Lagrangian—a theory that restricts its own field strength $b$—takes the form\footnote{See definitions of $F$ and $\tilde{G}$ just below, which in turn define the magnetic induction $\vec{B}$, the~electric field $\vec{E}$, and its ``dual'' magnetic field $\vec{H}$ and electric displacement $\vec{D}$, much like in Maxwell's theory. Also check Ref.~\cite{battesti-Rizzo(2016), battesti-rizzo(2013)} for a discussion on the limits to NLED and on the expanded B-I Lagrangian given next, in addition to~the magnetic and electric properties of a quantum vacuum.}
\bea
\label{B-I-lagrangian}
& &
L^{^{B-I}} (F, \tilde{G}) = - b^2 \left( \sqrt{ 1 + \left[ \frac{F}{2 b^2} - \frac{\tilde{G}^2}{16 b^4} \right] } -1 \right)\, ,
\\
& &
L^{^{B-I}}_{expd.} (F, \tilde{G})   \simeq \frac 1
2 F + \frac{1}{8b^2} F^2 + \frac{\tilde{G}^2}{2b^2} \,({\rm to\,fields\,lowest\,order }) \, .
\nonumber
\\
& &
{{It\,satisfies:}}
\nonumber
\\
& &
{\rm{(a)\,the\,canonical\, conjugate\,variable\,/\,constitutive\,relations\,}}
\nonumber
\\
& &
\vec{H} = - \frac{\partial L}{\partial \vec{B}} :::: \vec{D} = \frac{\partial L}{\partial \vec{E}} \, ,
\\
& &
{\rm{(b)\,the\,duality\,transformation}}
\nonumber
\\
& &
\vec{D} + i\,\vec{B} \rightarrow e^{i\theta}\,(\vec{D} + i\,\vec{B}) ::  
\vec{E} + i\,\vec{H} \rightarrow e^{i\theta}\,(\vec{E} + i\,\vec{H})\, ,
\\
& &
{\rm{and\, (c)\,\,the\,highly\,nontrivial\,and\,nonlinear\,constraint}}
\nonumber
\\
& &
\vec{D} \vec{H} =  \vec{E} \vec{B} \, .
\eea
In Equation~(\ref{B-I-lagrangian}), the term $b \equiv B^{^{B-I}}_{\rm lim}$ is an unknown parameter, the~theory's experimentally fixed field strength bound, with~the dimension of [mass$^2$] in natural units, that can be set using lab measurements---In~this instance, from ATLAS/LHC CERN. Hereafter, we shall use it as $b \equiv B^{B-I}_{ATLAS} \simeq \,3.28\cdot10^{22}\,{\rm G}$, 
the mass bound obtained in 
 \cite{Ellis-etal.(2017)}. The~corresponding electric field strength limit is  given as:

$$E^{^{B-I}}_{\rm lim} = \frac{e}{4\pi \epsilon_0 a_0^2} \simeq 2\cdot 10^{20} {\rm{\,[Volt/m]}},$$
where $a_{_0}$ = $\frac{4\pi \epsilon_0 \hbar^2}{\bar{m}_e e^2}$ = $\frac{\hbar}{\bar{m}_e c \alpha}$ = $5.29\cdot10^{-11}$\,m is the characteristic electron size, the~so-called Bohr radius~\cite{Born-Infeld(1934)}, and~$\bar{m}_e$ is the QED effective---one-loop corrected---electron mass \linebreak${\rm \bar{m}_e = m_e} \,\left[1+\frac{\alpha}{4 \pi}  \,ln^2\left(\frac{B^{B-I}_{lim}}{B_{Sch}}\right)\right]$ \cite{Feynman(QED-1985)}. 

In closing this paragraph, we define the invariants $(F, \tilde{G})$ (scalars) by using $F_{\mu\nu} (A_\mu) \equiv \nabla_\mu A_\nu - \nabla_\nu A_\mu \Longleftrightarrow F_{\mu\nu}(E_\mu, B_\nu) \equiv E_\mu V_\nu - E_\nu V_\mu + \eta_{\mu \nu}^{\hskip 0.3truecm \alpha \beta} V_\alpha B_\beta$, where the 4-vector $V^\alpha$ satisfies the normalization condition: ($V_\mu V^\mu$=1), so that the Maxwell scalar reads:  
$F \equiv F_{\mu\nu}\,F^{\mu\nu} = -2(\epsilon_{_0}{E}^2 - \mu_{_0}^{-1}{B}^2)$, and~the $F$-dual: $\tilde{G} \equiv F_{\mu\nu}{\bar{F}}^{\mu\nu} = \frac{1}{2} \eta_{\alpha \beta \gamma \delta} F^{\alpha \beta} F^{\gamma \delta} = \frac{1}{2}  \eta_{\mu \nu}^{\hskip 0.3truecm \alpha \beta} F_{\alpha \beta} F^{\mu\nu} = B_\mu E^\mu = \sqrt{\frac{ \epsilon_{_0} }{\mu_{_0}}}{\textbf E} \cdot {\textbf B}$, where 
the bi-vector ${\bar F}^{\mu\nu} \equiv \epsilon^{\mu\nu \rho\sigma} F_{\rho\sigma}$, with~the 4-tensor $\epsilon^{\alpha \beta\gamma \delta} \equiv \frac{1}{2 \sqrt{-g}}\;\varepsilon^{\alpha\beta \gamma\delta}$, being $\varepsilon^{ \alpha\beta \gamma\delta}$ the Levi-Civita tensor satisfying the indexing rule $\varepsilon_{_{0123}} = +1$. The~space--time metric signature (-+++) is used, along with~ the coordinate indexes $\mu,\nu$=$(0:t,1:r,2:\theta,3:\phi)$. Finally, $\epsilon_{_0} = 8.854 \cdot 10^{12}$\,[F/m]: vacuum permittivity; $\mu_{_0} = 4\pi \cdot 10^{-7}$\,[H/m]: vacuum~permeability.

On this basis, the~general equation of motion for nonlinear current/charge free electromagnetic field (EM) sources described by NLED Lagrangians ($L(F,\tilde{G})$) reads
\begin{equation}
\label{eqsofmotion}
\frac{1}{\sqrt{-g}} \nabla_\mu \left(\sqrt{-g} \left[- L_{F} F^{\mu\nu} - L_{\tilde{G}} {\bar{F}}^{\mu\nu} \right] \right) = 0 = J^\nu(\rho,{\bf J})\, ,
\end{equation}
where $\nabla_\mu$ is a covariant derivative with respect to the spherically symmetric Schwarzschild line-element (see below), $L_F = \partial L(F,\tilde{G}) /\partial F \propto \hbar, B^2/b^2$, and~$L_{\tilde{G}} = \partial L(F,\tilde{G})/\partial {\tilde{G}} = 0$ hereafter, and~ 
\be
L_{FF} = \partial^2 L(F,\tilde{G})/ \partial F^2 = \hbar \times (L_{FF}/L_{F}, B^2/b^2)\ldots  
\ee
Moreover, the~Faraday circularity law (Bianchi identity of first order) also applies
\begin{equation}
\label{bianchi} 
\nabla_\mu {\bar{F}}^{\nu \mu} \equiv \nabla_\mu F_{\nu\lambda} + \nabla_\nu F_{\lambda\mu} + \nabla_\lambda
F_{\mu\nu}=0\,. 
\end{equation}
It is needed to derive the effective metric Equation~(\ref{NLED-eff-metric}) presented below, by~using the notation introduced in~\cite{IJMPA(2006)}: $\nabla_\lambda F_{\mu\nu}|_\Sigma = f_{\mu\nu} k_\lambda \therefore f_{\mu\nu}$: field discontinuity at surface $\Sigma$, and~wave vector $k_\lambda$.

\subsubsection*{General 
 Lagrangians: Dispersion Relation and Effective~Metric}

For general NLED Lagrangians $L(F,\tilde{G})$, the~\textit{dispersion relation} that describes electromagnetic fields propagation gets the form ($g_{\mu \nu}$: background metric---Schwarzschild, Kerr, etc., 
 $k_{\nu}$: wave vector) \cite{novello-etal(2000), GRAV-REDSHIFT(2004), novello-etal(2000A), novello-etal(2001)}:
\begin{equation}
\underbrace{ \left(g_{\mu\nu} - 4 \frac{L_{FF}} {L_F + \tilde{G}\Omega{(F,\tilde{G})}_{\pm} L_{\tilde{G} \tilde{G}}} F_\mu \mbox{}^\lambda  F_{\nu}\mbox{}_{\lambda} \right) }_{ g_{\mu \nu}^{\rm eff} } 
k^{\mu} k^{\nu} = 0\; . 
\label{NLED-eff-metric}
\end{equation}

It was explicitly deduced in Appendix A in~\cite{IJMPA(2006)}.
This equation allows us to define the \textit{effective metric} (${ g_{\mu \nu}^{\rm eff} } $) that is "felt" at the gravitational field---``spacetime''---surrounding the remnant QGP core we study in this paper. Additionally, it was shown therein that the fields detected by a comoving observer in this NLED description of the average electromagnetic fields are given by 
$\langle F_{\alpha\mu} F^{\alpha}_{\,\,\,\nu} \rangle = - \frac{1}{3} B^2 h_{\mu\nu}$, where $B^2 \equiv B^\mu B_\mu $, in~addition to these other definitions: $B_\mu = |B|\,l_\mu , l_\mu = \sqrt{-g_{rr}}\,\delta^r_\mu , V_\mu = \sqrt{g_{00}}\, \delta^0_\mu$, with~$V_\mu V^\mu = 1$ representing the normalized 4-vector velocity of the local reference frame where the fields are~measured. 

NLED Lagrangians ($L(F, \tilde{G})$) have Planck's constant $\hbar$ as the expansion parameter on the field scalars ($F, \tilde{G}$); for~instance, the Heisenberg--Euler (H-E) NLED \cite{battesti-Rizzo(2016), battesti-rizzo(2013)}:
\be
L^{H-E}(F, \tilde{G}) = -\frac{1}{4}F + \frac{\tilde{\mu}}{4} \left(F^2 + \frac{7}{4} \tilde{G}^2 \right)\, ,
\label{Heisenberg-Euler-Lagrangian}
\ee
for which the following $\hbar$-dependent constant is defined: 
\be
\tilde{\mu}(\hbar) = \hbar\, \left(\frac{2 {\alpha}^2}{45}\right) \left\{ \frac{\hbar^2} {(m_e^4 c^5)} \right\} = \underline{A}\,\hbar \, \therefore \,  
{\alpha} = \frac{e^2} {(4\pi \hbar c)}\, . 
\ee

This feature classifies NLED as a semi-classical theory. Further, it is gauge-invariant; hence, charge conservation is guaranteed ($J^\nu_{\,\, ,\nu}$=$0$), while the theory vacuum is C-P-T symmetry-preserving. Moreover, NLED has the remarkable property of avoiding the singularity featuring the electric field at the position of a point-charge particle in Maxwell's theory. In addition, it fits well in regimes of extremely high magnetic field strengths, i.e.,~near to or beyond the Schwinger limit $\rm |B|_{_{\rm Sch}} \simeq 10^{13.5}$\,G, at~which quantum effects like vacuum polarization suddenly kick in, a~phenomenon that does not show up for any field strength, in accordance with Maxwell's~electrodynamics. 

Thus, by~resorting to Equation~(\ref{NLED-eff-metric}), the~effective metric for the H-E NLED Lagrangian ($L^{H-E}(F, \tilde{G}) \therefore \tilde{G}=0 $) becomes (recall that the H–E Lagrangian does not take into account all of the microscopic phenomena related to the photon--photon interaction in a vacuum).

\bea 
g_{\mu \nu}^{\rm eff} & = & g_{\mu \nu} \, {f\left(L^{\rm{H-E}} \left(\tilde{\mu}(\hbar) \, , \left\{\frac{B}{b}\right\}\right) \right)} 
\\ 
& = & g_{\mu \nu}  \left[ 1 - 
\underbrace{ \hbar \left(\frac{e^4}{90 \pi m_e^4 c^7}\right) }_{\tilde{\mu}(\hbar)} \left\{2 \mu_{_0}^{-1} \left(\frac{B^2}{b^2} \right) \right\}  + {\cal{O}} (\hbar^2) \ldots  \right]\, \nonumber 
\\ 
& = & g_{\mu \nu} \, f \left(\frac{B^2}{b^2} \right) (\textrm{unitless}),
\eea
where $B \equiv B^{Star}_{QGP}(G)$ takes on specific values near to/or well beyond the Schwinger limit, in~this QGP star~model. 

Meanwhile, and~after resorting once again to Equation~(\ref{NLED-eff-metric}), the~effective metric for the Born--Infeld NLED Lagrangian reads 
($L^{^{B-I}}(F, \tilde{G}) \therefore \tilde{G}=0$):\,\cite{IJMPA(2006)}
\bea 
& & 
g^{eff}_{\mu\nu} = g_{\mu\nu} - \frac{2B^2/b^2}{(2B^2/b^2 + 1)} \bigg\{V_\mu V_\nu - l_\mu l_\nu \bigg\} \,.
\\
& & \Longrightarrow \Longrightarrow  g^{eff}_{rr} = g_{rr} - \bigg\{ \frac{2B^2/b^2}{(2B^2/b^2 + 1)} \bigg\} \, g_{rr}
\\
& & 
g^{eff}_{\mu\nu} = g_{\mu\nu} \,
{\cal{F}} \left(\frac{B^2}{b^2} \right) (\textrm{unitless})
\, ,
\eea
as the $r$-metric component, which for extremely strong field strengths $B^{Star}_{QGP}(G) \lesssim B^{B-I}_{ATLAS}$ can mean a non-negligible modification of the Schwarzschild gravitational field geometry, in~this first approach to modeling a QGP star permeated by extremely high magnetic~fields.

To summarize, QED predicts that in most overarching settings involving high energy, photons can scatter off each other by exchanging (via Feynman diagrams) virtual charged fermions ($e^\pm$-like leptons and quarks$^\pm$) that acquire their masses through Yukawa-type interactions with the Higgs field. That is, the~Yukawa coupling constant couples the fermions---each having its own coupling constant---to either the Higgs field or the W$^{\pm}$ bosons. The~dynamics of such peculiar behavior of light was advanced decades ago (1951) by Karplus and Neuman~\cite{Karplus-Neuman(1951)}. Since then, lots of research on this subject has been carried out (see Dymnikova's review on nonlinear electrodynamics~\cite{Dymnikova(2021)}). In~the framework of NLED, Refs.~\cite{GRAV-REDSHIFT(2004), IJMPA(2006)} discuss the surface gravitational redshift and Einstein's gravitational lensing, essential concepts of relativistic astrophysics that help to justify why the QGP star presented below must be totally~dark.

\subsection{Einstein's Field Equations and QGP Star Dynamics Influenced by~NLED}

The functional action for nonlinear electrodynamics coupled minimally to gravity can be written as~\cite{Gravitation(1973), Blau(2020)}
\begin{equation}
\label{nled-action}
S = \int d^4x \, \sqrt{-g}\, \left(\frac{1}{2 \kappa} R + L_{_{M}} +  \left[  L_{_{NLED}} (F,\tilde{G}) \right] \right) \,.
\end{equation} 
The Einstein field equations are then written as 
\be
G_{\mu\nu} + \Lambda g_{\mu\nu} = R_{\mu\nu} - \frac{1}{2} g_{\mu\nu}R + \Lambda g_{\mu\nu} = \frac{8\pi G_{_{\rm N}} } {c^{4}} \big \langle 
T^{\rm{^{Total}} }_{\mu\nu} \big \rangle .
\label{Einstein-Eqs}
\ee

In the present study case, $\Lambda=0$. Thus, the~system energy--impulsion tensor reads
\be
\langle T^{\rm{^{Total}}}_{\mu\nu}\rangle = \langle T^{^{NL}}_{\mu\nu} \rangle + T^{^{M} }_{\mu\nu}, 
\ee
where the symbol $\langle \,\, \rangle$ stands for semi-classical (or ``renormalised'') stress--energy tensor. The~right-hand-side terms include the average contribution (VEV) from NLED electromagnetic fields and the perfect fluid thermal-like~matter. 

\subsubsection*{NLED 
and the Energy--Momentum Tensor of  Collapsing HHMNS~Cores}

After this lengthy digression, let us turn back to the model dynamics. First, the~energy--momentum tensor of an ordinary thermal (particle-like) fluid reads 
\be
{T}^{^M}_{\mu\nu} =  (\rho_{_M} c^2 + p_{_M}) v_\mu v_\nu - p_{_M} g_{\mu\nu}, 
\ee
with $\rho_{_M}$: mass density, and~$p_{_M}$: pressure. Here, the fluid 3-velocity field is $u$, so that the unit tangent vector field $v_\mu$ over the worldline of a particle reads 
\be
v_\mu=(-\frac{c}{\sqrt{c^2-u^2}},\frac{u}{\sqrt{c^2-u^2}})\,
\ee
which is properly squared-normalized such that $v_\mu v^\mu=-1$. This means that the 4-vector $v_\mu$ is associated with a time-like killing field $\zeta$ of the space--time~geometry.  

Second, keep in mind that when the above conditions are fulfilled, the~general NLED Lagrangian density $\{L_{_{NLED}}(F,\tilde{G}) \rightarrow L_{_{NL}}(F,\tilde{G})\}$ leads to an averaged energy--impulse tensor, also characterising the following for a~perfect-fluid 


\be
\langle T^{^{NL}}_{\mu\nu} \rangle = 
\langle (\rho_{_{NL}} + p_{_{NL}}) v_\mu v_\nu - p_{_{NL}} g_{\mu\nu} \rangle 
\label{nled-tmunu},
\ee
in which the NLED density and pressure are written like~\cite{novello-etal(2000A)} 
\bea
&& 
\rho_{_{NL}} = -L - 4 (\epsilon_{_0} E^2) L_F ,
p_{_{NL}} =  L + \frac 4 3 \big( \epsilon_{_0} E^2 - 2 \frac{B^2}{\mu_{_0}} \big) L_F\,.
\nonumber
\\
&& 
\therefore
\rho_{_{NL}} + 3p_{_{NL}} \leq 0 
\rightarrow L_F \geq \frac{L}{4 \mu_{_0}^{-1} B^2} 
\rightarrow \rightarrow \ddot{r} \geq 0\,\big\uparrow_{_{\rm{out}}}.
\eea

That is, the~QGP star fluid's effective acceleration is radially outward. This feature is dynamically contrary to the standard picture of gravitational collapse. Hence, the~total density ($\rho^{\rm Total}_{tt} \, \in T_{\mu\nu}$) and total pressure $(p^{\rm Total}_{ii} \in T_{\mu\nu}) \therefore \mu,\nu=(0:t,1:r,2:\theta,3:\phi)$ gather contributions from matter ($M$) and nonlinear EM fields ($NL$) to get the form\footnote{For a Maxwell Lagrangian the conserved Poynting flux energy density reads: 
$ \rho c^2 = 3 p = \frac{1}{2} (\epsilon_{_0} E^2 + \mu_{_0}^{-1} B^2) \doteq \sigma_{ij}$, the~Maxwell stress tensor.}
\begin{eqnarray}
&& \rho^{\rm Total}_{(r,B)} = \rho_{_{NL}} + \rho_{_{M}} = [-L -4 \left(\epsilon_{_0} E^2\right) L_F] + \rho_{_{M}} \, , 
\\
&& p^{\rm Total}_{(r,B)} = p_{_{NL}} + p_{_{M}} =  [L + \frac 4 3 \left(\epsilon_{_0} E^2 - 2 \mu_{_0}^{-1} B^2\right) L_F] + p_{_{M}} 
\, . 
\nonumber
\label{total-dens-press}
\end{eqnarray} 

Therefore, these equations indicate that NLED pressures and energy densities satisfy the relation $p_{_{NL}} = - \rho_{_{NL}}\,(c^2)$, as~discussed in~\cite{novello-etal(2000A), MPL-A(2010)} and references therein.
This NL equation of state (EoS) resembles the one of a fluid dominated by dark energy that drives accelerating cosmic expansion in the current standard model of cosmology (${ \Lambda}$-CDM: $\rho_{_{NL}} +  p_{_{NL}} = 0 =
- \frac{8}{3} \big(\epsilon_0 E^2 + \mu_{_0}^{-1} B^2 \big) L_F$. 
Thus, $L_F$\,=\,0 $\rightarrow$ the effective cosmological constant ${ \Lambda}$, i.e.,~{w\,=\,$-$1}), which in addition to the repulsive force among quarks derived from the QCD asymptotic freedom phenomenon (recall the quote about this mentioned above) generates a total repulsive pressure powerful enough to stably sustain the QGP star formed in this manner against the immense gravitational pull from its constitutive magnetized mass--energy. From~our understanding, the~behavior of quantum matter at such extreme pressures and field strengths is currently unknown and the subject of intensive research. The~role of both physical properties is the crux of the present~article.

 
\subsection{Extreme Magnetic Fields vs. Spherical Models and Actual Deformability of Ultra-Compact Star~Cores}

In this paper, when discussing the stability of an idealized spherical generic figure of equilibrium of a QGP star model---our first approach to tackling this issue\footnote{A realistic spinning QGP model is in progress~\cite{rotating QGP Star Model}.}---we shall explore magnetic field strengths in the~range

\be
({\rm 10^{18}\,<\,B^{Star}_{QGP}(G)\,<\,10^{22} }).
\ee

The last being the experimental bound computed from ATLAS/LHC data~\cite{Ellis-etal.(2017)}.

In connection with the highest value of magnetic field strengths achievable by astrophysical ultra-compact remnants like a quark star (see the recent review in~\cite{Recent-progresses-strange-qs}), it is essential to state that, at present, our understanding is unclear. That is, the~established properties of extremely hypermagnetized matter in new classes of compact stars might actually be quite different from what scientists previously thought. Despite foregoing spectroscopy tools, electromagnetic radiation satellites, optical and radio telescopes, and~the advent of GW astronomy, it is an extremely hard task to envision a physical mechanism to estimate those actual values at the inner core of an object like an HHMNS, let alone to attempt to infer them by direct observation of a unique compact remnant like a QGP star, of~which we  barely have observational evidence of its presence as an astrophysical body really lurking out there. 
This is the case of the candidate object RX J1856.5-3754, whose X-ray spectrum allows us to infer that it is consistent with a radius ($3.8 < {\rm{R^{X-R}_{\star}(km)}} < 8.2$) that is too small to be a typical neutron star. Even stranger is the remnant star at the center of the supernova debris HESS J1731-347, with~a radius of ${\rm{R^{HESS}_{\star}(km)}} \simeq 10.4$, a temperature of {$\rm{T(^oK)\sim 2\cdot10^6}$}, and~a mass of M${\rm{^{HESS}_{\star}(km)}} \simeq 77\%$\,M$_\odot$, certainly smaller than the Sun's mass. An~object with no indication of a thin over-layer---crust---composed of iron and/or other nuclei~\cite{Strange-Quark-Star(2022)}. Other possibilities are discussed in~\cite{Kate-Scholberg(2021)} and references therein. Hence, the~evidence in favor of QGP stars and their extreme magnetic fields is good (about seven strange-quark candidates are now known) but by no means conclusive. Hence, some innovative theoretical insights on how to compute or estimate NLED magnetic fields inside such quark stars are~needed.

For this reason, we quote the theoretical input by Usov and Shabad \cite{Usov-Shabad(2006)} regarding extreme magnetic fields, who computed a maximum B field strength: ${\rm B_{max}=1.6 \times 10^{28}}$\linebreak${\,B_{Sch} \sim 10^{42} }$\,G by imposing the stability condition for the existence of the ground state of the positronium atom. Namely, although~there is no experimental/observational evidence of such extreme terrific fields, we suggest that such a QED theoretical limit offers further support to our approach in building a model of a stable extremely hypermagnetized QGP star with much higher magnetic field strengths than the bounds computed in~\cite{Ellis-etal.(2017)} from the \mbox{ATLAS/LHC results:} 

\be
({\rm 10^{13.5}\,<\,B^{Star}_{QGP}(G)\,<\,10^{22\,--->\,42} }) .
\ee

These fields can bring about dramatic effects when computing astrophysical properties of the QGP remnant, such as, for~example, the~mass--radius relationship, as~given in Equation~(\ref{mass}) below. As it is such a critical property, we address it later~on. 

Hereafter, we assume this theoretical spherical configuration despite the actual nearly prolate/oblate/spheroidal  structure dictated by the effects of the powerful, essentially  dipolar magnetic (B) field permeating an actually rotating QGP star. (Recall what the observational properties of most typical neutron star/pulsars indicate.)
To this end, we highlight at this point the issue of anisotropic (polar vs. equatorial magnetic field) pressures\footnote{The magnetic pressure can be derived from the magneto-hydrodynamics Cauchy momentum equation: $\rho\,\left(\frac{\partial}{\partial t} + {\bf v\cdot\nabla} \right)\,{\bf v}= 
{\bf J}\times {\bf B} - 
{\bf \nabla} p$: (Lorentz Force - Pressure Force), where by using Ampere's law $\mu_0 {\bf J} = \nabla \times {\bf B}$, one gets 
${\bf J}\times {\bf B} = \frac{\left({\bf{B \cdot \nabla}}\right)\,{\bf B} }{\mu_0} - \nabla \left(\frac{B^2}{2\mu_0} \right)$:  (Magnetic Tension - Magnetic Pressure Force), with~$\mu_0=4.7\pi\times10^{-7}$\,W/A-m\,vacuum permeability.}, 
$p_{_{B}} = B^2/2\mu_0$, which were discussed in
~\cite{IJMPD(2007), Quant-magn-gravit-collapse, EPJ-C(2003)}. 

There it was shown that such pressure asymmetry---the superstrong field strength anisotropy---can induce a sudden quantum magnetic gravitational collapse to likely render a quark star as the astrophysical end state. In~fact, it has been suggested elsewhere that some observables of spheroidal magnetized strange stars can be estimated from a distance, for~example, by~detections of continuous gravitational wave signals from actually rotating and non-axisymmetric deformable QGP stars. See Refs.~\cite{papers-NS-deformation, abbott-etal(2022), haskell-etal(2008), tidal-deformation(2022),  Aurora-2021}, including the search for such a sort of GW signals by the LIGO/VIRGO/KAGRA~Collaboration.


At this point, it is essential to peruse the actual deformability of an ultra-compact star core in connection to our spherical model choice. Quite a lot of research has been carried out on this fundamental property, as it is deeply connected to the actual equation of state (EoS) of the compact object in question~\cite{LIGO-EOS_w_GW170817, Mass-Radius-NICER-Data}. See, for~instance, some studies on this matter in~\cite{papers-NS-deformation}. Ellipticity in the range of (0.48--0.93) $\times 10^{-9}$ (a few micrometres on the km scale radius) has been considered. In~addition, models in which the ratio of the polar to the equatorial radius reads $\sim$2~$\times~10^{-2}$, led to the conclusion that such a deformation level does not impede the consistent building of models of compact objects with a spherically symmetric structure (metric) \cite{papers-NS-deformation-1}. Similarly, other research hints at the existence of a trend in deformation relations, in~the case of hyper-strong fields, that are quasi-universal, i.e.,~mostly independent of the neutron star EoS~\cite{papers-NS-deformation-2}. In~view of these novel studies on deformable compact objects, we are confident that our model of a spherical QGP star enshrouded by a magnetic field of poloidal configuration is self-consistent, and~our conclusions are noteworthy. (A realistic general relativistic spinning model must certainly engender further novelties.) Indeed, in~our understanding, the~actual mechanical stiffness of the astrophysical QGP star is expected to be a little more stringent---i.e., less prone to undergoing noticeable deformation---than the one for typical neutron stars, and~thus its deformability is still quite limited in spite of the extremely strong magnetic field $B^{Star}_{QGP}$ acting on it. This is a fundamental physical difference in the actual structures of a neutron star and a QGP star. That is, it hides the fundamental role of the magnetic moment of quantum matter (electrons, protons, neutrons, etc.).

To conclude this summary, bear in mind that the whole Equation~(\ref{NLED-eff-metric}) also serves to describe the dispersion relation for light propagation in the gravitational field around compact remnants permeated by super-strong magnetic fields where NLED photon acceleration occurs. This effect is responsible for the asymptotically enlarging surface gravitational redshift $z_{_{\rm Grav}}$ of compact stars, a~fundamental piece of relativistic astrophysics in this paper. The~dramatic effect of such extreme field strengths on this crucial astrophysical quantity is illustrated in Figure \ref{REDSHIFT-1-EFFETIVO} (see~\cite{GRAV-REDSHIFT(2004), mendonca-book, IJMPA(2006), mendonca-etal(2006)}).  

{Einstein once said that
``\ldots  it is theory which decides what is `observable'.''}, 
and 
{John S. Bell asserted: 
``\ldots I think he
was right---'observation' is a complicated and theory-laden business.''}

\begin{figure}[h!]
\includegraphics[width=8.0cm]{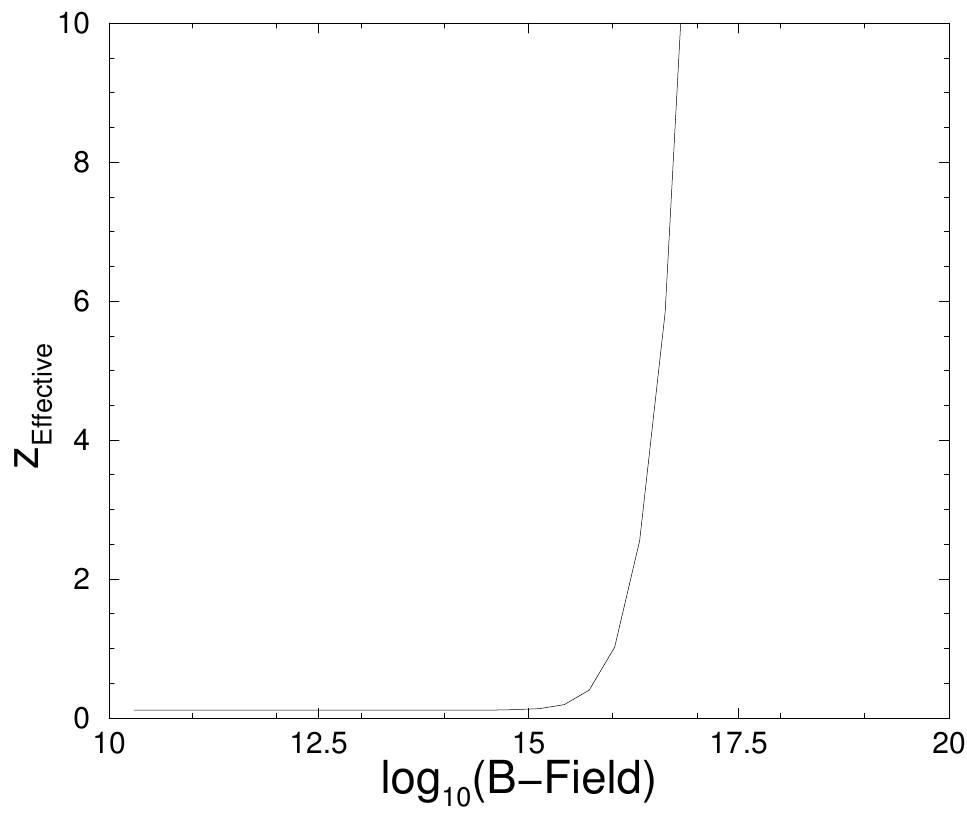}
\caption{Effective surface gravitational redshift $z_{_{\rm Grav}}$ for B-fields up to $10^{20}$\,G. This graph clearly indicates the trend of an increase in $z_{_{\rm Grav}}$ for the extremely large B-fields, as the subject of this analysis. Its trend is exponentially increasing but it never reaches an infinite value, as~is the case for true BHs. (Plot taken from the paper \cite{IJMPA(2006)}). } 
\label{REDSHIFT-1-EFFETIVO}
\end{figure}



{Conceptualisation, Methodology, Calculations and Analysis, H.J.M.C., Methodology and Calculations; F.H.Z.G., W.D.A.P. and E.M.S., Methodology and Analysis; G.U.A.F. and R.F.L. All authors have read and agreed to the published version of the manuscript.}

{This research received no external funding }


{No data are available.}

\acknowledgments{The authors extend their heartfelt gratitude to the numerous scientists who contributed to leading this work to finally see the light. It has been a lengthy journey. Indeed, we appreciate all the referees' valuable criticisms and suggestions, as they helped to put forward the central arguments supporting the key goal of the present article.  
Special thanks to Jorge E. Horvath (IAG-USP/São Paulo/SP, Brazil) and Odylio D. de Aguiar (INPE/São Jos\'e dos Campos/SP, Brazil) for their very first reading and criticisms of this paper. 
HJMC specially thanks the Rodrigo Francisco dos Santos (Idealizer of ''Aether Tenebris Channel'' on YouTube) and Luis Gustavo de Almeida (Department of Physics,  Universidade Federal do Acre, Rio Branco, Acre,  Brazil) co-authors of the alternative version of the original idea addressed in this paper. 
HJMC also thanks the ''Mr. Pacho'' Internet site chief C{\'e}sar A. Chavarria and the LAN Operator Durany Laverde for their hospitality and assistance with system issues during the preparation of this document in Medellin, Colombia.}

{The authors declare no conflicts of interest. The funders had no role in the design of the study; in the collection, analyses, or interpretation of data; in the writing of the manuscript; or in the decision to publish the results.}

\clearpage 

\end{document}